\documentclass[
twocolumn]{aastex631}
\usepackage{amsmath}

\begin{document}

\title{Exploring the anisotropic gravitational wave background from all-sky mock gravitational wave event catalogs}

\correspondingauthor{Xi-Long Fan}
\email{E-mail: xilong.fan@whu.edu.cn}

\author{Zhencheng Li}
\affil{National Astronomical Observatories, Chinese Academy of Sciences, Beijing 100101, China}
\affil{School of Astronomy and Space Sciences, University of Chinese Academy of Sciences, Beijing, 100049, China}

\author{Zhen Jiang}
\affil{National Astronomical Observatories, Chinese Academy of Sciences, Beijing 100101, China}

\author{Yun Liu}
\affil{National Astronomical Observatories, Chinese Academy of Sciences, Beijing 100101, China}
\affil{School of Astronomy and Space Sciences, University of Chinese Academy of Sciences, Beijing, 100049, China}

\author{Xi-Long Fan}
\affil{School of Physics and Technology, Wuhan University, Wuhan 430072, China}

\author{Liang Gao}
\affil{School of Physics and Astronomy, Beijing Normal University, Beijing, 100875, China}
\affil{School of Physics and Microelectronics, Zhengzhou University, Zhengzhou 450001, China}

\author{Yun Chen}
\affil{National Astronomical Observatories, Chinese Academy of Sciences, Beijing 100101, China}
\affil{School of Astronomy and Space Sciences, University of Chinese Academy of Sciences, Beijing, 100049, China}

\author{Tengpeng Xu}
\affil{National Astronomical Observatories, Chinese Academy of Sciences, Beijing 100101, China}
\affil{School of Astronomy and Space Sciences, University of Chinese Academy of Sciences, Beijing, 100049, China}

\begin{abstract}
Anisotropic stochastic gravitational wave background (SGWB) serves as a potential probe of the large-scale structure (LSS) of the universe. In this work, we explore the anisotropic SGWB from local ($z < \sim 0.085$) merging stellar mass compact binaries, specifically focusing on merging stellar binary black holes, merging neutron star–black hole binaries, and merging binary neutron stars. The analysis employs seven all-sky mock lightcone gravitational wave event catalogs, which are derived from the Millennium simulation combined with a semianalytic model of galaxy formation and a binary population synthesis model. We calculate the angular power spectra $\mathrm{C}_\ell$ at multipole moments $\ell$, expressed as $\text{log}_{10} [\ell(\ell+1)\mathrm{C}_\ell/(2\pi)]$, based on the skymaps of the overdensity $\delta_\mathrm{GW}$ in the anisotropic SGWB. The spectra for all three source types exhibit an approximately linear increase with $\text{log}_{10} \ell$ at higher $\ell$ (e.g., $\ell > \sim 30 - 300$) in seven catalogs, with a characteristic slope of $\sim 2$. The spectra of seven catalogs exhibit considerable variations, arising from fluctuations in spatial distribution, primarily in the radial distribution, of nearby sources (e.g., $< 50$ Mpc/h). After subtracting these nearby sources, the variations become much smaller and the spectra for the three source types become closely aligned (within discrepancies of a factor of $\sim 2$ across $\ell = 1 - 1000$ for all catalogs). We also find that including farther sources results in a rapid decrease in the anisotropy.
\end{abstract}


\section{Introduction} 

Various gravitational wave (GW) sources, such as merging supermassive ($M>10^7M_\odot$) binary black holes (SMBBHs), merging stellar binary black holes (BBHs), merging neutron star–black hole binaries (NSBHs), merging binary neutron stars (BNSs), contribute to the total stochastic gravitational wave background (SGWB). For a summary of the sources, see recent reviews, e.g., \citet{2019RPPh...82a6903C} and \citet{2022Galax..10...34R}. Recently, pulsar timing array collaborations, including the European Pulsar Timing Array \citep{2023A&A...678A..50E}, NANOGrav \citep{2023ApJ...951L...8A}, Parkes Pulsar Timing Array \citep{2023ApJ...951L...6R}, and Chinese Pulsar Timing Array \citep{2023RAA....23g5024X}, have provided strong evidence for the existence of the SGWB. Furthermore, direct detections of merging BBHs, merging BNSs, and merging NSBHs \citep[e.g.,][]{2023PhRvX..13d1039A} also suggest the presence of a potentially vast number of compact binary coalescences (CBCs) contributing to the SGWB.  The anticipated detection of the SGWB by both ground-based detectors, such as the advanced Laser Interferometer Gravitational-Wave Observatory (aLIGO) \citep{2010CQGra..27h4006H} and the Einstein Telescope (ET) \citep{2010CQGra..27s4002P}, and space-based detectors, such as the Laser Interferometer Space Antenna (LISA) \citep{2017arXiv170200786A}, is forthcoming, promising significant advancements in our understanding of the universe.

The SGWB can be divided into isotropic and anisotropic parts. Beyond the extensively studied isotropic part \citep[e.g.,][]{2021PhRvD.104b2004A, 2022A&A...660A..26B, 2023ApJ...952L..37A}, the anisotropic part provides additional insights, particularly concerning the spatial distribution of GW sources. Several studies have explored the anisotropic SGWB; for example, the anisotropy from merging SMBBHs \citep{2017NatAs...1..886M, 2022ApJ...941..119B, 2023ApJ...956L...3A, 2024ApJ...965..164G,2024arXiv240805043Y}, and the anisotropy from merging BBHs, NSBHs, BNSs \citep{2018PhRvL.120w1101C, 2018PhRvD..98f3501J, 2019PhRvL.122k1101J, 2021JCAP...11..032C}. \citet{2018PhRvD..98f3501J} and \citet{2019PhRvL.122k1101J} first use the mock lightcone GW event catalog to investigate the anisotropy of merging stellar-mass compact binaries. Specifically, they employ a postprocessing analytical approach to compute the GW events within the galaxies contained in the all-sky mock lightcone galaxy catalog sourced from \citet{2005MNRAS.360..159B} and \citet{2007MNRAS.375....2D}. However, their results suggest a significantly higher anisotropy level compared to the prediction of \citet{2018PhRvL.120w1101C}, particularly at very small scales. The discrepancies among these studies remain under discussion. 

The GW event catalogs could be obtained from a semianalytic model of galaxy formation, Galaxy Assembly with Binary Evolution (GABE) \citep{2019RAA....19..151J}, combined with a rapid binary population synthesis model COSMIC v3.3.0 \citep{2020ApJ...898...71B, katie_breivik_2020_3905335}. Using the binary population data generated by COSMIC, merging BNSs, NSBHs and BBHs can be explored within the cosmological framework provided by GABE. For example, \citet{2024MNRAS.527.5616L} calculates the total SGWB from these sources using this GABE-COSMIC methodology. However, the original event catalog lacks information on the source positions within the celestial sphere. In this work, we use the original event catalog to construct seven all-sky mock lightcone GW event catalogs, which incorporate source position information across the celestial sphere, to investigate the anisotropic SGWB from the local ($z < \sim 0.085$) merging BBHs, NSBHs, and BNSs.

The structure of this paper is as follows. In Section~\ref{sec:Formalism}, we present the fundamental equations used to calculate the anisotropic SGWB and the angular power spectrum. Section~\ref{sec:Simulation of sources} briefly outlines the GABE-COSMIC methodology to produce the GW event catalogs and details the methodology employed for generating mock lightcone GW event catalogs. The results, including the anisotropic SGWB and angular power spectra, are discussed in Section~\ref{sec:Results}. Finally, the primary conclusions of the study are summarized in Section~\ref{sec:Summary and Discussion}.

\section{Formalism}
\label{sec:Formalism}

\subsection{Anisotropic SGWB from ground-based detector compact binaries}
\label{subsec:Ground-based detectors sources}

The SGWB from the observed direction $\hat{e}$ is defined by the dimensionless energy density \citep{2017PhRvD..96j3019C, 2018PhRvD..98f3501J}:
\begin{equation}
    \Omega _\mathrm{GW}\left ( f , \hat{e} \right )= \frac{1}{\rho _\mathrm{c}}\frac{\mathrm{d}^3 \rho _\mathrm{GW}\left ( f, \hat{e} \right )}{\mathrm{d} \mathrm{ln}\left ( f \right )\mathrm{d}^2\Omega},
	\label{eq:Omega origin}
\end{equation}
where ${\rho_\mathrm{c}}=3{{H_0}^2}{c^2}/(8{\pi}G)$ is the critical energy density and ${\mathrm{d}^3\rho_\mathrm{GW}}$ is the present-day energy density of GWs from the observed frequency interval $\{f, f+\mathrm{d}f\}$ and the solid angle element $\mathrm{d}^2 \Omega$ centered on $\hat{e}$. The $\Omega_\mathrm{GW}$ could be calculated using the Phinney formula \citep{2001astro.ph..8028P} in the homogeneous and isotropic universe:
\begin{equation}
\begin{aligned}
\Omega_\mathrm{GW}(f,\hat{e})=&\dfrac{1}{4 \pi \rho_c }\int \int\frac{\mathcal{N}(z,\hat{e},\vec\gamma)}{1+z}  \left [\frac{\mathrm{d}E (f_r,\vec\gamma)}{\mathrm{d}\ln (f_r)}\right]_{f_\mathrm{r}} \,dz \, \mathrm{d}\vec\gamma ,
	\label{eq:Omega Phinney}
\end{aligned}
\end{equation}
where $\vec{\gamma}$ are intrinsic parameters of GW sources, such as the primary star mass $m_1$ and the secondary star mass $m_2$ of CBCs in this work; $\mathcal{N}(z, \hat{e},\vec{\gamma}) =\mathrm{d}^3N(z, \hat{e},\vec{\gamma})/(\mathrm{d}V_{\mathrm{c}, \mathrm{e}} \mathrm{d}z\mathrm{d}\vec\gamma)$ is the number of GW sources per comoving volume per redshift per intrinsic parameter range at observed direction $\hat{e}$; and $\mathrm{d} E\left (f_r, \vec{\gamma} \right )/\mathrm{d ln}\left ( f_r \right )$ is the energy spectrum of a single GW source in logarithmic interval, $f$ is the observed frequency, and $f_\mathrm{r} = (1+z)f$ is the frequency in the cosmic rest frame of the GW source. In this work, we adopt the same energy spectrum as used in \citet{2011ApJ...739...86Z} and \citet{2008PhRvD..77j4017A}, assuming that this spectrum remains applicable for merging BNSs and NSBHs. By dividing the full-direction comoving element presented in \citet{2001astro.ph..8028P} by $4\pi$, we can derive the comoving volume element at the observed direction $\hat{e}$ as follows:
\begin{equation}
    \mathrm{d}V_{\mathrm{c}, \mathrm{e}} = \frac{c\, d^2_\mathrm{c}}{H(z)}\mathrm{d}z,
	\label{eq:comoving volume element}
\end{equation}
where $H(z) = H_0 [\Omega _\mathrm{m}\left (1+z\right )^{3}+\Omega _\Lambda ]^{0.5}$ and $d_\mathrm{c}$ is the comoving distance:
\begin{equation}
        \label{eq:d_c}
        \mathrm{d}_\mathrm{c}(z) = \int_{0}^{z} \frac{c\mathrm{d}z'}{H_0\sqrt{\Omega_\mathrm{m}(1+z')^3+\Omega_\Lambda } }. 
    \end{equation}
Using $\mathrm{d}z/\mathrm{d}t_\mathrm{r} = (1+z) H(z)$, $\mathcal{N}(z,\hat{e},\vec{\gamma})$ could be calculated using the merger rate of GW sources, defined by the number of mergers per source-frame time $t_\mathrm{r}$, in the comoving volume element in the intrinsic parameter element, $R(z, \hat{e}, \vec{\gamma}) = \mathrm{d}N(z, \hat{e},\vec{\gamma})/\mathrm{d}t_{r}$, at observed direction $\hat{e}$:
\begin{equation}
    \mathcal{N}(z, \hat{e},\vec{\gamma}) = \frac{\mathrm{d}^3 N(z, \hat{e},\vec{\gamma})}{c(1+z)d^2_\mathrm{c} \, \mathrm{d}t_r \, \mathrm{d}z\, \mathrm{d}\vec\gamma}= \frac{\mathrm{d}^2 R(z, \hat{e},\vec{\gamma})}{c(1+z)d^2_\mathrm{c} \,\mathrm{d}z\, \mathrm{d}\vec\gamma}.
	\label{eq:N and R}
\end{equation}
We now extend the applicability of the above formulas to an inhomogeneous and anisotropic universe, assuming that all GW events occur within galaxies. Consequently, $\mathcal{N}(z, \hat{e},\vec{\gamma})$ and $\Omega _\mathrm{GW}\left (f, \hat{e} \right )$ might vary across different directions in accordance with the distribution of galaxies. By substituting Eq. \ref{eq:N and R} into Eq. \ref{eq:Omega Phinney}, we compute the direction-dependent SGWB by summing the contributions from all galaxies within the galaxy catalog (similar to \citealt{2018PhRvD..98f3501J}):
\begin{equation}
\begin{aligned}
    \Omega _\mathrm{GW}\left ( f,  \hat{e}\right ) &= \frac{f}{4\pi\mathrm{c}\rho _\mathrm{c}}\sum^{k=k_\mathrm{tot}}_{k=1}\sum^{m = m_\mathrm{tot}}_{m=1}\sum^{n = n_\mathrm{tot}}_{n=1} \\
    & \frac{\mathrm{R}_k(z_k, m_{1, m}, m_{2,n})}{(1+z_k)d^2_{\mathrm{c}, k}} \delta_\mathrm{K} (\hat{e}, \hat{e}_k) \\
    & \times \left [ \frac{\mathrm{d} E\left ( f_\mathrm{r}, m_{1\mathrm{cen}, m}, m_{2\mathrm{cen}, n} \right )}{\mathrm{d}  f_\mathrm{r}} \right ]_{f_\mathrm{r}=\left ( 1+z_k \right )f} \\
    &= \sum^{k=k_\mathrm{tot}}_{k=1}\Omega _{\mathrm{GW}, \mathrm{gal}, k} (f) \delta_\mathrm{K} (\hat{e}, \hat{e}_k),
\end{aligned}
\label{eq:Omega in use}
\end{equation}
where $m$ and $n$ are the bin indices of the primary star mass $m_1$ and the secondary star mass $m_2$ of CBCs; $m_{1\mathrm{cen}}$ and $m_{2\mathrm{cen}}$ are the central values of $m_1$ and $m_2$ in each bin; and $m_\mathrm{tot}$ and $n_\mathrm{tot}$ are the total bin numbers of $m_1$ and $m_2$, respectively, which are determined by the division of the $m_1 \times m_2$ space; $k$ is the index of the galaxies; $k_\mathrm{tot}$ is the total number of the galaxies in the catalog; $\mathrm{R}_k$ is the merger rate of merging stellar compact binaries in the galaxy k at $z_k$ in the primary mass-secondary mass bin $\Delta m_1, m_2$; $\delta_{K}$ is the Kronecker delta function; $\hat{e}_k$ is the direction of the galaxy $k$, and see Sec. \ref{subsec:Mock construction} for details on the positions of our galaxies; and $\Omega _{\mathrm{GW}, \mathrm{gal}, k}$ is the contribution of the galaxy $k$ to the SGWB. In this work, the $\mathrm{R}_k$ is approximated as the mean merger rate in the galaxy $k$, denoted $\overline{\mathrm{R}}_k$. Details regarding the calculation of $\overline{\mathrm{R}}_k$ can be found in Sec. \ref{sec:Simulation of sources}. This study focuses exclusively on the anisotropy arising from the spatial distribution of GW sources. For a more comprehensive formalism that incorporates additional effects such as Doppler, Kaiser, and gravitational potential influences, see \citet{2017PhRvD..96j3019C} and \citet{2020PhRvD.101j3513B}. Note that some of these effects have a negligible impact on anisotropy. For example, the kinematic dipole resulting from the observer's peculiar motion has been reported as minor \citep{2018PhRvD..98f3501J, 2022arXiv220205105J}.

In this work, the skymaps are constructed as HEALPix\footnote{http://healpix.sf.net} maps \citep{2005ApJ...622..759G} composed of $\mathrm{N}_{\mathrm{pix}}$ equal area pixels, where $\mathrm{N}_{\mathrm{pix}}=12\times\mathrm{N}_{\mathrm{side}}^2$ is the number of pixels of the HEALPix skymap, with $\mathrm{N}_{\mathrm{side}}$ denoting the parameter utilized in the HEALPix framework. A pixel may correspond to a range of directions, specifically $\hat{e}_{i, 1}$, $\hat{e}_{i, 2}$, ..., $\hat{e}_{i, n_i}$. In practice, $\Omega_{\mathrm{GW}}(f, \hat{e}_i)$ is computed as follows:
\begin{equation}
\begin{aligned}
    \Omega _\mathrm{GW}\left ( f,  \hat{e}_i\right )= \frac{1}{\Delta\Omega}\sum^{n=n_i}_{n=1}\sum^{k=k_\mathrm{tot}}_{k=1}\Omega _{\mathrm{GW}, \mathrm{gal}, k} (f) \delta_\mathrm{K} (\hat{e}_{i, n}, \hat{e}_k),
\end{aligned}
\label{eq:Omega in use2}
\end{equation}
where $i$ denotes the index of the pixel corresponding to the observed direction $\hat{e}_i$ ($\hat{e}$ associated with the $i$th pixel), $n_i$ is the total number of the directions within the pixel, and $\Delta\Omega$ is the pixel area.

The overdensity $\delta_\mathrm{GW}$ in the anisotropic SGWB from ground-based detector compact binaries is defined as follows \citep{2019PhRvL.122k1101J}:
\begin{equation}
    \delta _\mathrm{GW}\left ( f,  \hat{e} \right )= \frac{\Omega _\mathrm{GW}\left ( f,  \hat{e} \right ) - \overline{\Omega} _\mathrm{GW}\left ( f \right )}{\overline{\Omega} _\mathrm{GW}\left ( f \right )},
	\label{eq:Omega overdensity}
\end{equation}
where $\overline{\Omega} _\mathrm{GW}\left ( f \right ) = \frac{1}{4\pi}\int_{\mathcal{S}^2}\Omega _\mathrm{GW}\left ( f,  \hat{e} \right )\mathrm{d}^2\Omega$ is the mean SGWB over the sky and $\mathcal{S}^2$ is the surface of the celestial sphere. 

\subsection{Angular power spectrum}
\label{subsec: Angular power spectrum}
To quantify the anisotropy of the SGWB at various scales, the angular power spectrum $\mathrm{C}_\ell$ is adopted in this work. The angular power spectrum $\mathrm{C}_\ell$ quantifies the amplitude of statistical fluctuations in the angular power of the SGWB at scales $\theta \sim 180^{\circ}/\ell$, where $\ell$ denotes the multipole moment. The \( \mathrm{C}_\ell \) is calculated using the healpy package \citep{2005ApJ...622..759G, Zonca2019}. Below are its formulas \citep{gorski2024healpixprimer}. The overdensity $\delta_\mathrm{GW}$ at each frequency on the celestial sphere can be expanded as:
\begin{equation}
            \label{eq:the decomposed}
            \delta_\mathrm{GW}(\hat{e})= \sum_{\ell, m}\mathrm{a}_{\ell m}\mathrm{Y}_{\ell m}\left ( \hat{e}  \right ),
        \end{equation}
where $\mathrm{Y}_{lm}$ are spherical harmonics and:
\begin{equation}
            \label{eq:clm}
            \mathrm{a}_{\ell m} = \frac{4\pi }{\mathrm{N}_{\mathrm{pix}} } \sum_{i=0}^{\mathrm{N}_{\mathrm{pix}}-1} \mathrm{Y}_{\ell m}^\ast (\hat{e}_i ) \, \delta_\mathrm{GW}(\hat{e}_i ),
        \end{equation}
where $i$ denotes the index of the pixel, and the superscript star is complex conjugation. Consequently, the angular power spectrum can be expressed as:
\begin{equation}
            \label{eq:Cl}
            \mathrm{C}_\ell  = \frac{1}{2\ell+1}\sum_{m = -\ell}^{m = +\ell} \left |  \mathrm{a}_{\ell m} \right |^2.
        \end{equation}
We calculate $\mathrm{C}_\ell$ up to $\ell_{\mathrm{max}}=3\mathrm{N}_{\mathrm{side}}-1$\footnote{See https://healpix.sourceforge.io/html/fac\_anafast.htm.}. To maintain consistency with the previous literature, we multiply $\mathrm{C}_\ell$ by $\ell(\ell+1)/(2\pi)$. Note that using the angular power spectrum alone to describe the anisotropy implicitly assumes a stationary Gaussian random field for the SGWB. However, when considering the SGWB at a single frequency, along with the effective temporal observation window for ground-based GW detectors, additional non-Gaussianity might arise \citep{2018PhRvD..98f3501J, 2018PhRvD..98f3509J}. In future work, bispectrum and trispectrum analyses might be required to capture these complexities in greater detail.

\section{Simulation of sources}
\label{sec:Simulation of sources}

\subsection{The GW events in the GABE-COSMIC methodology}
\label{subsec:GABE model}

GABE is a semianalytic model of galaxy formation, specifically addressing the evolution of binaries. This model is based on the Millennium simulation \citep{2005Natur.435..629S}, which has a simulation box volume of $(500/h)^3 \mathrm{Mpc}^3$. In this simulation, a fiducial cosmological model with $\Omega_\mathrm{m}=0.25$, $\Omega_\Lambda=0.75$ and $H_0=73 \mathrm{km}/\mathrm{s}/\mathrm{Mpc}$ \citep[WMAP1,][]{2003ApJS..148..175S} is adopted. Further details on the generation of merging compact binaries using the GABE-COSMIC methodology can be found in \citet{2024MNRAS.527.5616L}, and we utilize datasets identical to those described therein. Here, we provide a concise overview. We denote $t_\mathrm{r}$ as $t$ hereafter.

We begin by discussing the formation of merging stellar-mass compact binaries. Within the GABE framework, the total stellar populations in galaxies result from star formation events. Each star formation event triggers the birth of a simple stellar population (SSP) under conditions such as an unstable gaseous disk or galaxy mergers. The number, mass distribution, and delay time distribution of merging stellar-mass compact binaries originating from each SSP are determined using the rapid binary population synthesis model COSMIC. The final population of merging stellar-mass compact binaries within galaxies is a composite of these individual SSP contributions. 

Upon executing GABE, we record for each galaxy the redshift, evolution history, spatial coordinates within the simulation box ($x_0$, $x_1$, $x_2$), mean merger rates $\overline{\mathrm{R}}_k(z_k, m_1, m_2)$, and the mass distribution of the merging BNSs, NSBHs, and BBHs contained within each galaxy, across 64 snapshots at discrete redshifts. The selection of the discrete redshifts is aligned with the Millennium simulation. The mean merger rate $\overline{\mathrm{R}}_k = \Delta N_k/ \Delta t$, where $\Delta t$ represents the cosmic time interval between the current snapshot of galaxy $k$ and its preceding snapshot, and $\Delta N_k$ denotes the number of merging sources of interest during $\Delta t$ within the galaxy $k$.

Compared to the modeling in \citet{2018PhRvD..98f3501J}, our GABE+COSMIC approach (see our previous work \citealt{2024MNRAS.527.5616L} for details) presents several potential advantages. First, our approach employs the binary population synthesis model COSMIC to simulate binary evolution and calculate the merger time delays and mass distribution of CBCs, offering a more physically motivated framework. For instance, it is capable of theoretically predicting the event rates of CBCs. Second, our approach dynamically calculates the CBC population of galaxies during the runtime of GABE, directly outputting the results to the snapshots. Third, GABE is based on several updated descriptions of galaxy formation processes (see \citealt{2019RAA....19..151J} and \citealt{2011MNRAS.413..101G} for details), including gas cooling, supernova feedback, angular momentum transfer, ejecta reincorporation, and satellite stripping, relative to \citet{2007MNRAS.375....2D}.

\subsection{All-sky mock lightcone GW event catalog construction}
\label{subsec:Mock construction}
\begin{figure}
    \centering
	\includegraphics[width=\columnwidth]{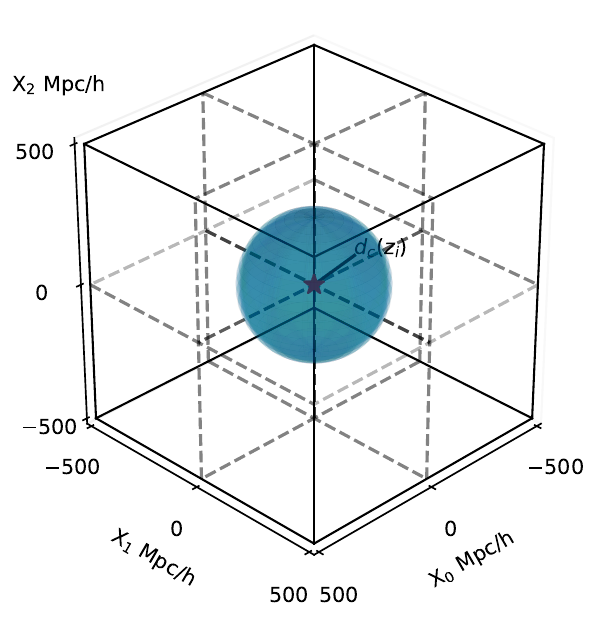}
    \caption{The new box, generated by replicating the initial simulation box eight times. The observer is positioned at the center, specifically at coordinates (0, 0, 0) Mpc/h, with the surface of the past-lightcone at redshift $z = z_i$ presented for reference.} 
    \label{fig:fig1}
\end{figure}

Using the GW event catalog derived from the GABE-COSMIC methodology, we create mock lightcone GW event catalogs restricted to the local universe, specifically for $ d_c< 250 \mathrm{Mpc/h}$ ($z < \sim 0.085$). Our methodology is detailed below. The initial simulation box is replicated eight times, with the observer positioned at the center of the resultant new box, as illustrated in Fig. \ref{fig:fig1}. The coordinates of objects within the new box are derived from the relation ($X_0$, $X_1$, $X_2$) = ($x_0-a$L, $x_1-b$L, $x_2-c$L), where $a$, $b$ and $c$ can take values of $0$ or $1$ and $\mathrm{L}=500\mathrm{Mpc/h}$.
The surface of the past-lightcone of the observer, representing the observable mock universe, at $z = z_i$, corresponds to the spherical surface with the radius equal to the comoving distance $\mathrm{d}_\mathrm{c}(z_i)$. 

A galaxy crosses the surface of the past-lightcone of the observer when its distance to the observer, given by $r = \sqrt{X_0^2+X_1^2+X_2^2}$, matches the comoving distance that light can travel from $z=0$ to the redshift of the target galaxy, denoted as $\mathrm{d}_\mathrm{c}(z)$. Since galaxies in the original catalog from
the GABE-COSMIC methodology are stored in discrete snapshots, the galaxy, along with its associated GW events, is included in the mock lightcone catalog if it meets two criteria: (1) in the initial snapshot at $z_\mathrm{ori}$, $r_\mathrm{ori}<\mathrm{d}_\mathrm{c}(z_\mathrm{ori})$, and (2) in the subsequent snapshot at $z_\mathrm{des}$, $r_\mathrm{des}>\mathrm{d}_\mathrm{c}(z_\mathrm{des})$.

To determine the exact time $t(z_\mathrm{obs})$ of the galaxy as it crosses the surface of the past-lightcone between $z_\mathrm{ori}$ and $z_\mathrm{des}$, two assumptions are employed: (1) the galaxy follows a straight-line trajectory between two adjacent snapshots\footnote{This assumption is widely used \citep[e.g.,][]{2005MNRAS.360..159B, 2007MNRAS.376....2K, 2013MNRAS.429..556M, 2019ApJS..245...26K}. Different interpolation methods for galaxy positions between snapshots show nearly identical precision, with small differences (e.g., $\sim 10^{-2}$ Mpc/$h$), as detailed in \citet{2022MNRAS.516.4529S}, that are negligible for our analysis.}, and (2) the comoving distance could be approximated to increase linearly with cosmic time $t(z)$ during the interval. These assumptions yield the following two equations:
\begin{equation}
\left\{\begin{aligned}
&\frac{t(z_\mathrm{obs})-t(z_\mathrm{ori})}{t(z_\mathrm{des})-t(z_\mathrm{ori})} = \frac{r_\mathrm{obs}-r_\mathrm{ori}}{r_\mathrm{des}-r_\mathrm{ori}} \\ 
&\frac{t(z_\mathrm{obs})-t(z_\mathrm{ori})}{t(z_\mathrm{des})-t(z_\mathrm{ori})} = \frac{\mathrm{d}_\mathrm{c}(z_\mathrm{obs})-\mathrm{d}_\mathrm{c}(z_\mathrm{ori})}{\mathrm{d}_\mathrm{c}(z_\mathrm{des})-\mathrm{d}_\mathrm{c}(z_\mathrm{ori})}
\end{aligned}\right..
\label{eq:pot}
\end{equation}
Consequently, $t(z_\mathrm{obs})$ corresponds to the time at which the two equations intersect, with the condition that at this intersection $r_\mathrm{obs} = \mathrm{d}_\mathrm{c}(z_\mathrm{obs})$. The position of the galaxy and the mean merger rates of merging stellar-mass compact binaries within the galaxy could be determined through linear interpolation:
\begin{equation}
    X_{i, \mathrm{obs}} = X_{i, \mathrm{ori}}+\frac{t(z_\mathrm{obs})-t(z_\mathrm{ori})}{t(z_\mathrm{des})-t(z_\mathrm{ori})}\times(X_{i, \mathrm{des}}-X_{i, \mathrm{ori}}),
\label{eq:interp position}
\end{equation}

\begin{equation}
\begin{aligned}
\overline{\mathrm{R}}_{\mathrm{obs}}(z_{\mathrm{obs}}, m_1, m_2) &= \overline{\mathrm{R}}_{\mathrm{ori}}(z_{\mathrm{ori}}, m_1, m_2)\\
& +\frac{t(z_\mathrm{obs})-t(z_\mathrm{ori})}{t(z_\mathrm{des})-t(z_\mathrm{ori})}\times 
    (\overline{\mathrm{R}}_{\mathrm{des}}(z_{\mathrm{des}}, m_1, m_2) \\
    &-\overline{\mathrm{R}}_{\mathrm{ori}}(z_{\mathrm{ori}}, m_1, m_2)),
\label{eq:interp R}
\end{aligned}
\end{equation}
where $i = 0, 1, 2$. The R.A. and decl. of the galaxy, which denote its directional orientation, are determined under the assumption that the coordinate system  of the observer aligns with the equatorial coordinate system.
By applying the aforementioned procedure to all galaxies with  distances to the observer less than $250 \mathrm{Mpc/h}$ in the original catalog from the GABE-COSMIC methodology, we construct the all-sky mock lightcone GW event catalog. We designate the all-sky mock lightcone GW event catalog, constructed with the observer positioned at coordinates (0, 0, 0) Mpc/h as described above, as the ``r1" model. In addition to the r1 model, we change the observer's location to generate six additional mock catalogs, labeled from ``r2" to ``r7" models. The coordinates of the observers for these models are detailed in Table. \ref{tab:modelr1r7}. We select the r2 model as our fiducial model because it demonstrates the most significant anisotropy in the merging BBHs among the seven models (see results in Sec. \ref{subsec:The angular power spectra}).
\begin{table}
	\centering
	\caption{Observer coordinates in the new box for the models ``r2" through ``r7", in addition to the observer coordinate (0, 0, 0) Mpc/h for the model ``r1"} 
	\label{tab:modelr1r7}
\begin{tabular}{|c|c|c|c|c|}
  \hline
  Model & r2 & r3 & r4\\
  \hline
  Pos/(Mpc/h) & (125, 0, 0) & (250, 0, 0) & (250, 125, 0)\\
  \hline
  Model & r5 & r6 & r7
  \\
  \hline
  Pos/(Mpc/h) & (250, 250, 0) & (250, 250, 125) & (250, 250, 250)
  \\
  \hline
\end{tabular}

\end{table}

The local merger rates from our seven catalogs are in good agreement with the results ($436.8 , 3.0 , 37.0$ $\mathrm{Gpc}^{-3} \mathrm{yr}^{-1}$ for merging BNSs, NSBHs, and BBHs, respectively) from our previous work \citet{2024MNRAS.527.5616L}, with differences within $\sim 30\%$. The local merger rates for BNSs and BBHs are also in strong agreement with the estimated local merger rates from GWTC-3 \citep{2023PhRvX..13a1048A}: $10$–$1700$ $\mathrm{Gpc}^{-3} \mathrm{yr}^{-1}$ for BNSs and $17.9$–$44$ $\mathrm{Gpc}^{-3} \mathrm{yr}^{-1}$ for BBHs. For NSBHs, the local merger rates in our study inherit the limitations from our previous work \citet{2024MNRAS.527.5616L} and are lower than the estimated local merger rates from GWTC-3. A relevant discussion can be found in the second paragraph of Section 4.1 in \citet{2024MNRAS.527.5616L}. However, using the overdensity to calculate the angular power spectrum would mitigate the underestimation. Furthermore, we did not observe any anomalous results for NSBHs when compared to BNSs and BBHs (see Section 4).

In principle, mock lightcone catalogs extending to higher redshifts could be constructed using a similar approach. However, due to computational limitations arising from the large number of galaxies and the associated merger rate information, our catalogs are restricted to $z \sim 0.085$. Certain strategies could mitigate the computational burden, such as applying selection criteria to filter out less massive galaxies. For comparison, the mock lightcone catalog from \citet{2005MNRAS.360..159B} used in \citet{2018PhRvD..98f3501J}, which extends to $z \sim 0.78$, contains only $\sim 42\%$ of the number of galaxies compared with ours because its galaxies were selected based on an apparent AB magnitude cut of $< 18$ in the r filter from Sloan Digital Sky Survey (SDSS).

\section{Results}
\label{sec:Results}

\subsection{The angular power spectra}
\label{subsec:The angular power spectra}
\begin{figure*}
    \centering
	\includegraphics[width=2.2\columnwidth]{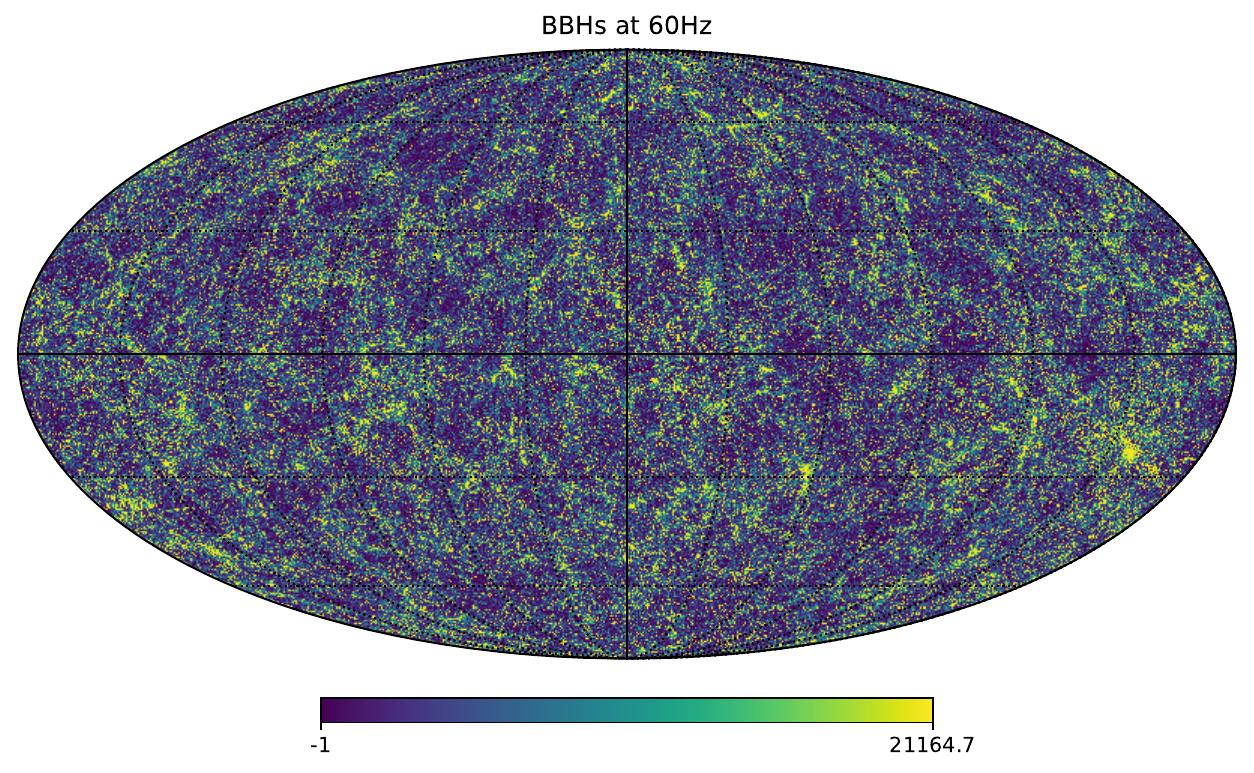}
    \caption{The HEALPix skymap of the overdensity $\delta_\mathrm{GW}$ in the anisotropic SGWB from the merging BBHs up to $d_\mathrm{c} = 250\mathrm{Mpc}/h$ at $60$ Hz with $N_\mathrm{side}=512$ for the r2 model.}
    \label{fig:BBH GABE distribution}
\end{figure*}

The HEALPix skymap of the overdensity $\delta_\mathrm{GW}$ in the anisotropic SGWB from the merging BBHs up to $d_\mathrm{c} = 250\mathrm{Mpc}/h$ at $60$ Hz with $N_\mathrm{side}=512$ for the r2 model is shown in Fig. \ref{fig:BBH GABE distribution}. The overdensity distinctly traces the large-scale structure (LSS) of the universe, highlighting the filaments, nodes, and voids of the cosmic web. The skymaps for merging BNSs and NSBHs display similar characteristics, which are not presented here.

\begin{figure}
	\centering
	\includegraphics[width=\columnwidth]{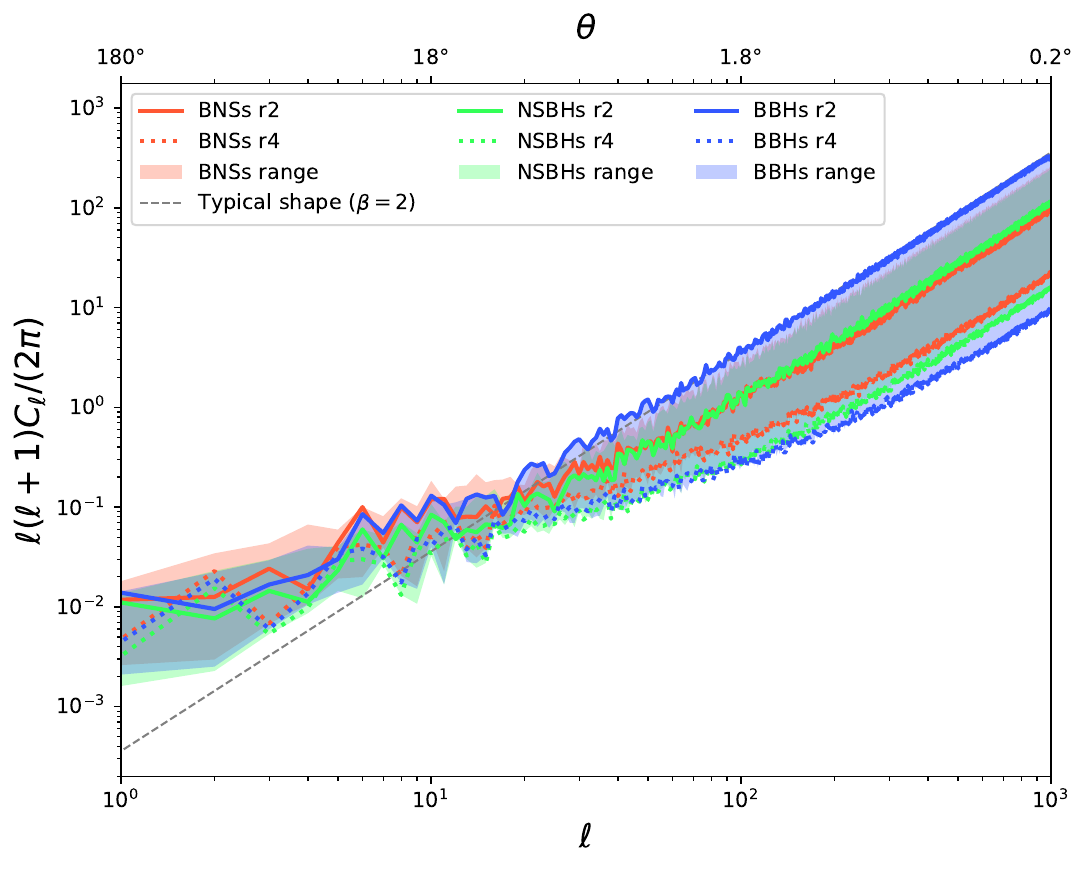}
    \includegraphics[width=\columnwidth]{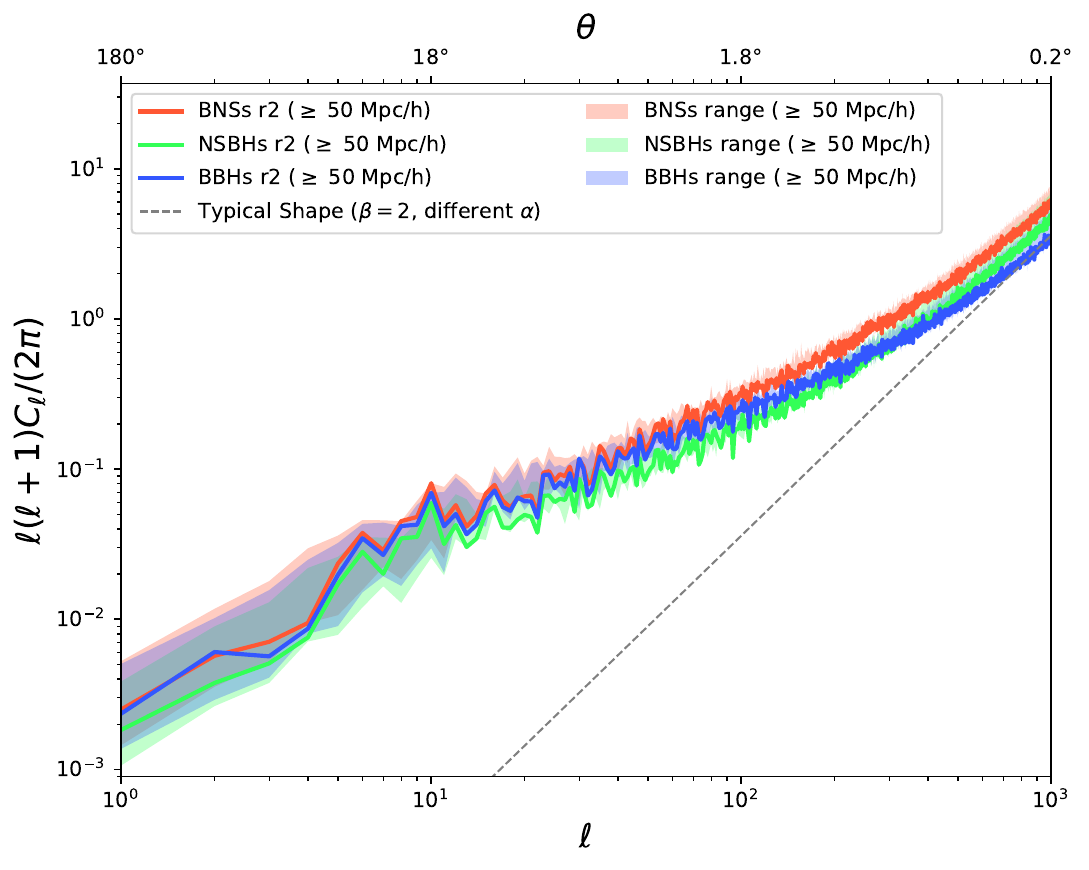}
	\quad
    \caption{Upper panel: The angular power spectra of the overdensity $\delta_\mathrm{GW}$ in the SGWB from merging BNSs, NSBHs and BBHs up to $d_\mathrm{c} = 250\mathrm{Mpc}/h$ at $60$ Hz respectively, with $N_\mathrm{side}=512$, for the seven models, ranging from r1 to r7. All the spectra are contained within the colored regions, with the spectra for models r2 and r4 distinctly highlighted. Lower panel: Similar to the upper panel, but with nearby ($ < 50$ Mpc/h) sources subtracted and exclusively highlights the spectra for the r2 model. The approximate scale of $\ell$ is indicated on the upper axis (rounded to one decimal place), along with reference lines (gray dashed lines) exhibiting slopes of $\beta = 2$ (note that the intercepts $\alpha$ are variable) in both panels.}
    \label{fig:cl}
\end{figure}

The upper panel of Fig. \ref{fig:cl} presents the angular power spectra of the overdensity $\delta_\mathrm{GW}$ in the SGWB from merging BNSs, NSBHs and BBHs up to $d_\mathrm{c} = 250\mathrm{Mpc}/h$ at $60$ Hz, respectively, with $N_\mathrm{side}=512$, for the seven models, ranging from r1 to r7. All the spectra are contained within the colored regions, with the spectra for models r2 and r4 distinctly highlighted.
The spectra for all three source types exhibit relatively flat shapes below $\ell \sim 10$, followed by steep upward trends.  For higher $\ell$ (e.g., $\ell > \sim 30 - 300$), the trends are especially typical, with slopes $\beta \approx 2$ in logarithmic space. The spectra across our seven catalogs exhibit considerable variations. For example, the spectra for merging BBHs exhibit a variation spanning approximately a factor of $2$ to one order of magnitude for $\ell = 1$ to $\sim 80$ and about $1$–$2$ orders of magnitude for $\ell = \sim 80$ to $1000$. Our results are not directly comparable to those of \citet{2018PhRvD..98f3501J} and \citet{2019PhRvL.122k1101J} for several reasons. First, the mock catalog used by \citet{2018PhRvD..98f3501J} extends to z $\sim 0.78$, significantly surpassing the redshift range of the catalog in our study. Consequently, the spectra derived from closer sources may be higher, as discussed later. Additionally, their catalog includes only galaxies with apparent AB magnitudes limited to $18$ in the r filter from SDSS, whereas our catalog encompasses all galaxies. However, we present a comparison here to highlight certain commonalities. Our spectra exhibit shapes similar to those reported by \citet{2018PhRvD..98f3501J} and \citet{2019PhRvL.122k1101J}  for higher $\ell$ (e.g., $\ell > \sim 30 - 300$). Their spectra for merging BBHs \footnote{The spectrum reported by \citet{2018PhRvD..98f3501J} includes all three types of merging stellar-mass compact binaries but introduces only a $\mathcal{O}(1)$ difference, as noted by the authors.} are generally situated near the lower boundary of the spectrum range for merging BBHs we obtained. We also conduct a brief comparison with the findings of \citet{2018PhRvL.120w1101C}, although our results are not directly comparable for several reasons. In addition to the limited redshift range explored in our study, differences in the GW energy spectrum and the models of merging stellar-mass compact binaries might also contribute to the disparity. The lower boundary of our spectra for merging BBHs exhibits significantly higher amplitudes ($>2$ orders of magnitude for the majority of $\ell$) compared to the spectrum in \citet{2018PhRvL.120w1101C}.\footnote{In our study, the comparison is actually made with the reproduced spectrum of \citet{2018PhRvL.120w1101C}, presented in Fig. 2 of \citet{2019PhRvL.122k1101J} as the ``CMBquick + Halofit" model at $65.75$ Hz.}

The distribution of several nearby sources might contribute significantly to the fluctuations in anisotropy. To illustrate this effect, the angular power spectra of the overdensity $\delta_\mathrm{GW}$ in the SGWB from merging BNSs, NSBHs and BBHs between $ 50$ and $250$  Mpc/h at $60$ Hz are shown, respectively, in the lower panel of Fig. \ref{fig:cl}. After the subtraction of the nearby sources, the growth of the spectra relatively slows down within the range $\sim 10 < \ell < \sim 600$, yet they continue to exhibit the typical sharp shapes  for $ \ell> \sim 600$. The variation in the spectra for the r1 to r7 models is significantly reduced across $\ell = 1 - 1000$. The spectra for all three merging types exhibit a variation of less than a factor of $\sim 4$ for $\ell=1$ to $\sim 20$, and less than a factor of $\sim 2$ for $\ell=\sim 20$ to $1000$. The sharp reduction in variation after the subtraction indicates that the significant discrepancies observed in the spectra of the upper panel of Fig. \ref{fig:cl} are primarily attributable to the nearby sources. For each of seven mock catalogs, approximately $10^5$ galaxies are situated within $50$ Mpc/h, accounting for roughly $0.5\%-1\%$ of the total galaxy population. However, the SGWB scales as $\mathrm{d}_\mathrm{c}^{-2}$, as indicated by Eq.\ref{eq:Omega in use}. Consequently, nearer sources contribute more significantly to the SGWB. The variation regions of the spectra after the subtraction of nearby sources exhibit greater similarity to the error regions described in \citet{2018PhRvD..98f3501J} and \citet{2019PhRvL.122k1101J} for higher $\ell$ (e.g., $\ell> \sim 20$). However, the shapes of the spectra more closely resemble the results reported by \citet{2018PhRvL.120w1101C} within the range  $\sim 10 < \ell < \sim 100$.  Note that to enable a more reasonable comparison with the existing literature in the future work, in addition to using a larger simulation box, several other factors should be addressed. For example, when comparing with \citet{2019PhRvL.122k1101J}, it is essential to apply a magnitude filter to our galaxy catalogs. Furthermore, understanding the influence of the GW energy spectra and the models of merging stellar-mass compact binaries on anisotropy is essential for comparisons with \citet{2018PhRvL.120w1101C}.

From the observational perspective, the contribution to the total SGWB from nearby sources would be the first to be subtracted by future GW detectors because these discrete events are the most readily detectable. As more nearby sources are resolved with accurate positional information, the significant uncertainty in the theoretically predicted angular power spectra of the overdensity in the SGWB, arising from the mapping of these sources onto the skymap, will progressively decrease. Ultimately, the theoretical angular power spectra for the residual SGWB will exhibit a much narrower uncertainty range, as shown in the lower panel of Figure. 3. Consequently, these spectra will be better suited for comparison with other theoretical predictions or future observational results, enabling the potential extraction of astrophysical (or cosmological) information, such as LSS. Subtracting nearby sources would also mitigate the necessity for highly accurate mock GW event catalogs in predicting the angular power spectra. After the subtraction of the nearby sources, the spectra for merging BNSs, NSBHs and BBHs at $60$ Hz display notable similarities (within discrepancies of a factor of $\sim 2$ across $\ell = 1 - 1000$ for all catalogs). This result suggests that the anisotropy is primarily influenced by the distribution of GW source hosts and might serve as a valuable probe of the LSS of the universe. Typically, anisotropy from CBCs can also be investigated through directly detected GW sources, as demonstrated in \citet{2020PhRvD.102j2004P}. However, studies of the SGWB offer complementary methods. In this study, our spectra establish a theoretical upper limit \footnote{Our analysis considers sources within the distance of $250$ Mpc/h. Incorporating more distant sources would reduce the anisotropy, as further discussed in Section 4.3.} on the anisotropy from CBCs. The spectra might also be compared with other angular power spectra defined by overdensity, such as the CMB spectra, in the future, potentially uncovering additional and deeper astrophysical and cosmological insights, such as those related to matter distribution and galaxy clusters. Additionally, an accurate angular power spectrum for the SGWB from astrophysical GW sources in the future could establish a limit for the detectable anisotropy of the SGWB originating from potential cosmological GW sources that cannot be individually detected, such as cosmic strings \citep[e.g.,][]{2018PhRvD..98f3509J}. Below this limit, these SGWB components might be masked.

\subsection{The effects of the radial distribution}
\label{subsec:The radial anisotropy}
The spatial distribution of the sources could be decomposed into two parts: the angular distribution on the celestial sphere, referred to as ``azimuthal distribution", and the distribution at different comoving distances, referred to as ``radial distribution". The influence of the radial distribution on the anisotropy is assessed through the implementation of the following smoothing process in our work. We partition the comoving distances of galaxies, along with the merging stellar-mass compact binaries they contain, into $m$  equally spaced bins within the range [$a$, $b$] Mpc/h, referred to as ``$m$-bin approximation". Here, $a$ and $b$ denote the lower and upper boundaries of the comoving distances of the considered sources, respectively. For example, in the $5$-bin approximation, the bin boundaries are defined as [$a$, $a+\frac{(b-a)}{5}$, $a+\frac{2(b-a)}{5}$, $a+\frac{3(b-a)}{5}$, $a+\frac{4(b-a)}{5}$, $b$] Mpc/h. The comoving distances of the galaxies within each bin, denoted as $\mathrm{d}_{\mathrm{c}, i, 1}$, $\mathrm{d}_{\mathrm{c}, i, 2}$ ... $\mathrm{d}_{\mathrm{c}, i, n_i}$ are then replaced by the mean comoving distance for that bin, given by $\overline{\mathrm{d}}_{\mathrm{c}, i}=\sum^{n=n_i}_{n=1} \mathrm{d}_{\mathrm{c}, i, n}/n_i$, where $i$ represents the bin index and $n_i$ is the total number of the galaxies within the bin. The variation in the spectra between those calculated directly and those obtained using the $m$-bin approximation is interpreted as the anisotropy potentially influenced by the radial distribution. We anticipate that the extreme case of $m=1$, corresponding to the $1$-bin approximation, captures the anisotropy  approximately arising from the azimuthal distribution. In this work, we use the approximation model to study sources up to $d_\mathrm{c} = 50\mathrm{Mpc}/h$ ($a=0, b=50$), between $50 - 250\mathrm{Mpc}/h$ ($a=50, b=250$) and up to $d_\mathrm{c} = 250\mathrm{Mpc}/h$ ($a=0, b=250$).

\begin{figure}
	\centering
	\includegraphics[width=\columnwidth]{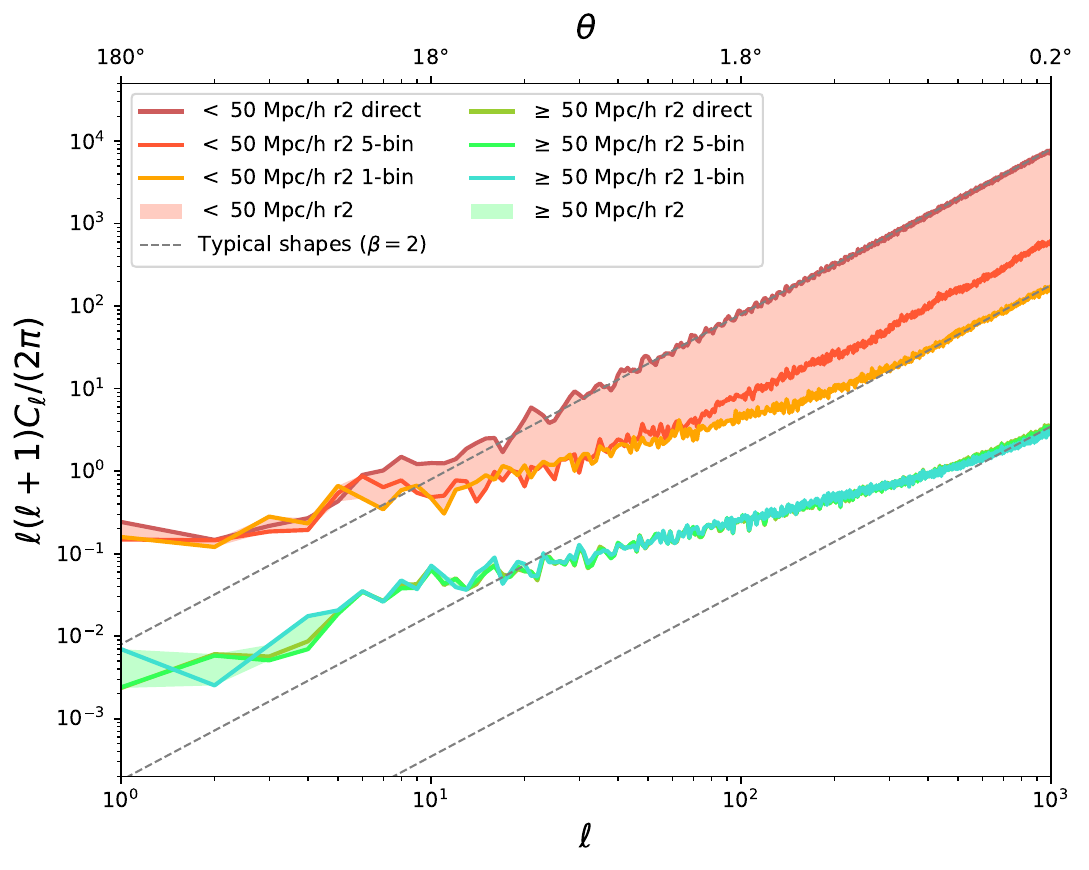}
    \includegraphics[width=\columnwidth]{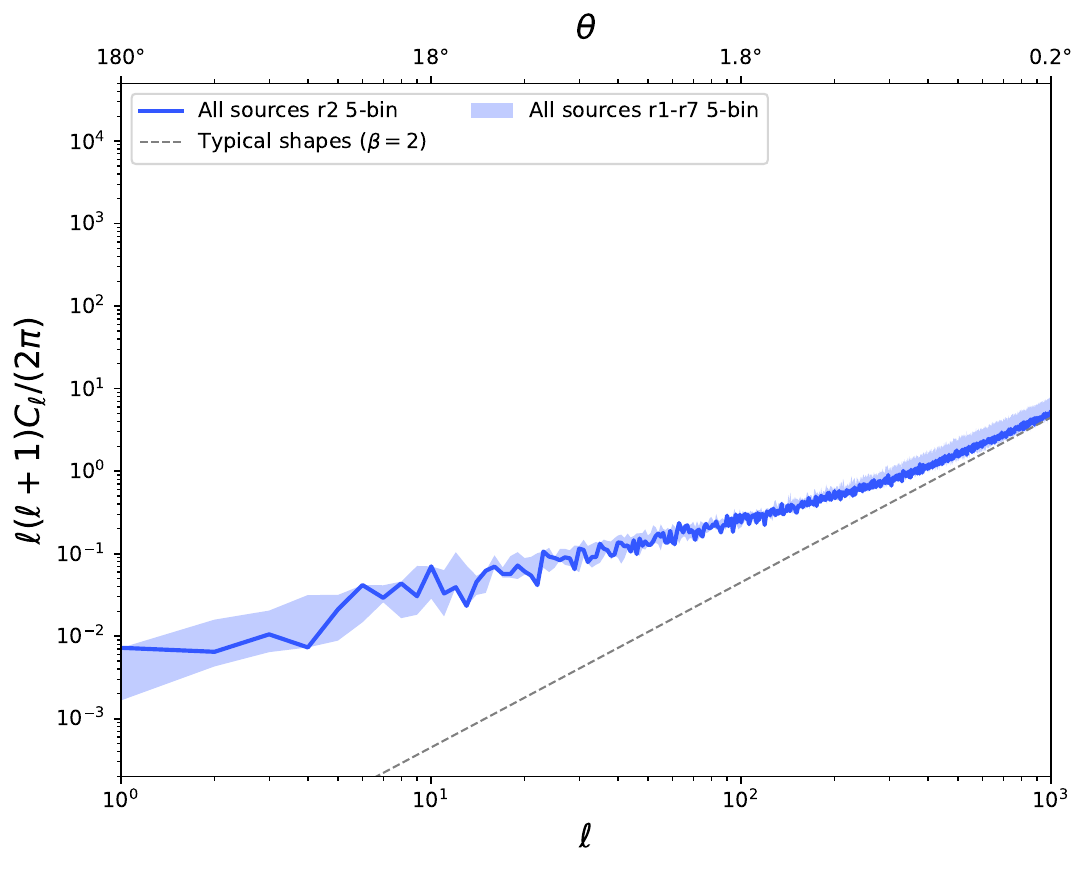}
    \caption{ Upper panel: The angular power spectra of the overdensity $\delta_\mathrm{GW}$ in the SGWB from merging BBHs within $d_\mathrm{c} = 50\mathrm{Mpc}/h$ and between $d_\mathrm{c} = 50 - 250\mathrm{Mpc}/h$, derived from direct calculation, the 5-bin approximation model, and the 1-bin approximation model, respectively, with $N_\mathrm{side}=512$ at $60$ Hz in the r2 model. The colored regions illustrate the influence of the radial distribution.  Lower panel: The angular power spectra of the overdensity $\delta_\mathrm{GW}$ in the SGWB from merging BBHs up to $d_\mathrm{c} = 250\mathrm{Mpc}/h$ at $60$ Hz, derived from the 5-bin approximation model, with $N_\mathrm{side}=512$, for the seven models, ranging from r1 to r7. All the spectra are contained within the colored regions, with the spectrum for the r2 model distinctly highlighted. The approximate scale of $\ell$ is indicated on the upper axis (rounded to one decimal place), along with reference lines (gray dashed lines) exhibiting slopes of $\beta = 2$.}
    \label{fig:clbins}
\end{figure}

To evaluate the impact of the radial distribution of both nearby and farther sources on the anisotropy, we calculate the angular power spectra of the overdensity $\delta_\mathrm{GW}$ in the SGWB from merging BBHs up to $d_\mathrm{c} = 50\mathrm{Mpc}/h$ and between $50 - 250\mathrm{Mpc}/h$, derived from direct calculation, the 5-bin approximation model, and the 1-bin approximation model, respectively, with $N_\mathrm{side}=512$ at $60$ Hz in the r2 model. The results are presented in the upper panel of Fig. \ref{fig:clbins}. The colored regions illustrate the influence of the radial distribution.
The variation of the spectra for the sources within $d_\mathrm{c} = 50\mathrm{Mpc}/h$ is within a factor of $\sim 3$ for $\ell = 1$ to $\sim 10$, a factor of $\sim 2$ to approximately one order of magnitude for $\ell = \sim 10$ to $\sim 50$, and $\sim 1-2$ orders of magnitude for $\ell = \sim 50$ to $1000$. For the sources situated between $50 - 250\mathrm{Mpc}/h$, the variation is within a factor of $\sim 3$ for $\ell = 1$ to $\sim 10$, and within about $30\%$ for $\ell = \sim 10$ to $1000$. Consequently, when modeling the anisotropy with the subtraction of nearby sources, the radial distribution of sources becomes less significant for higher $\ell$ (e.g., $\ell>\sim 10$). 
The directly calculated spectrum for sources up to $d_\mathrm{c} = 50\mathrm{Mpc}/h$ exhibits the typical upward shape for $\ell > \sim 30$, as shown by the ``BBHs r2" spectrum in the upper panel of Fig. \ref{fig:cl}. However, the directly calculated spectrum for sources within the range of $50 - 250\mathrm{Mpc}/h$ does not display this shape until $\ell > \sim 600$, as shown by the spectra in the lower panel of Fig. \ref{fig:cl}. Since observational factors are not incorporated in our analysis, the angular power spectra presented in our current work do not include instrumental noise. For further discussion on the angular power spectrum of the instrumental noise, see \citet{2020PhRvD.101l4048A}.

The lower panel of Fig.\ref{fig:clbins} shows the angular power spectra of the overdensity $\delta_\mathrm{GW}$ in the SGWB from merging BBHs up to $d_\mathrm{c} = 250\mathrm{Mpc}/h$ at $60$ Hz,  derived from the $5$-bin approximation, with $N_\mathrm{side}=512$, for the seven models, ranging from r1 to r7. All the spectra are contained within the blue regions, with the spectrum for the r2 model distinctly highlighted. The spectra exhibit minimal variation (less than a factor of $\sim 3.5$) across our seven mock catalogs for $\ell>\sim 10$. This observation also suggests that the significant variation in the spectra presented in the upper panel of Fig. \ref{fig:cl} for $\ell>\sim 30$ might primarily arise from fluctuations in the radial distribution of the nearby sources. The discrete nature of the processes underlying the SGWB, such as the discrete spatial and temporal distribution of GW sources, often leads to Poisson noise, also known as shot noise (see \citealt{2019PhRvD.100f3508J, 2020PhRvD.102b3002A, 2019PhRvD.100h3501J, 2024PhRvD.109j3535K} for further discussion). The fluctuations in the radial distribution in our analysis might mainly result from inherent Poisson noise in our catalogs, arising from the discreteness of galaxies (and the GW sources within them) in the radial distribution. The primary term contributing to the radial dependence in Eq. \ref{eq:Omega in use} , $\mathrm{d}_\mathrm{c}^{-2}$, originates from the calculation of the comoving number density $\mathcal{N}$ in Eq. \ref{eq:N and R}. Consequently, the radial distribution of sources represents the number density of galaxies within each comoving volume element $\mathrm{d}V_{\mathrm{c}, \mathrm{e}}$ at various redshifts, which inherently includes Poisson noise. When the number of galaxies is small (e.g. $\sim 10^5$ within 50 Mpc/$h$ in our models), this noise might predominate in the estimation of the anisotropic SGWB. 

We note that the spectra for both merging BBHs up to $d_\mathrm{c} = 50\mathrm{Mpc}/h$ and between $50$ and $250\mathrm{Mpc}/h$ exhibit typical shapes for higher $\ell$ (e.g. $\ell > \sim 300 -600$), even under the extreme 1-bin approximation. Consequently, other factors might contribute to the Poisson noise. The typical shapes of our spectra exhibit slopes similar to those of the modeled Poisson noise spectra presented in \citet{2019PhRvD.100f3004C}. Additionally, the spectral behavior aligns with the predictions made in \citet{2019PhRvD.100f3004C}, which suggest that the high $ \ell $ (e.g., $\ell > \sim 50$ to $\sim 600$, based on the different choice of integral cut-off, shown in their Fig. 9.) portion of the angular power spectrum is predominantly influenced by Poisson noise arising from the azimuthal distribution of mock catalogs. This might indicate that the Poisson noise from the azimuthal distribution is entangled within our spectra. In addition to Poisson noise, cosmic variance, arising from the variation in observer positions across our seven models, might also contribute to the fluctuations in the galaxy distributions within our catalogs. This might account for some of the variation in the angular power spectra across our seven models, particularly at lower values of $\ell$ (e.g., $\ell < \sim 10$). However, it is unlikely to be the primary factor directly contributing to the observed large variation of the angular power spectra for $\ell > \sim 10$ in the upper panels of Figs.\ref{fig:cl} and \ref{fig:clbins} because it primarily affects the angular power spectra at lower $\ell$ (e.g., $\ell < \sim 10$) and has minimal impact for higher $\ell$ (e.g., $\ell > \sim 10$), as shown in Fig. 4 of \citet{2018PhRvD..98f3501J}. Nevertheless, cosmic variance could have an indirectly profound impact on the angular power spectra because it reflects underlying variations in the distribution and number density of galaxies, which in turn influence the Poisson noise. For instance, in more densely populated regions of the universe, relative Poisson noise might be smaller, while in more sparsely populated regions, it might be larger.

\subsection{The effects of including farther sources}
\label{subsec:The effects of including farther sources}
\begin{figure}
	\centering
	\includegraphics[width=\columnwidth]{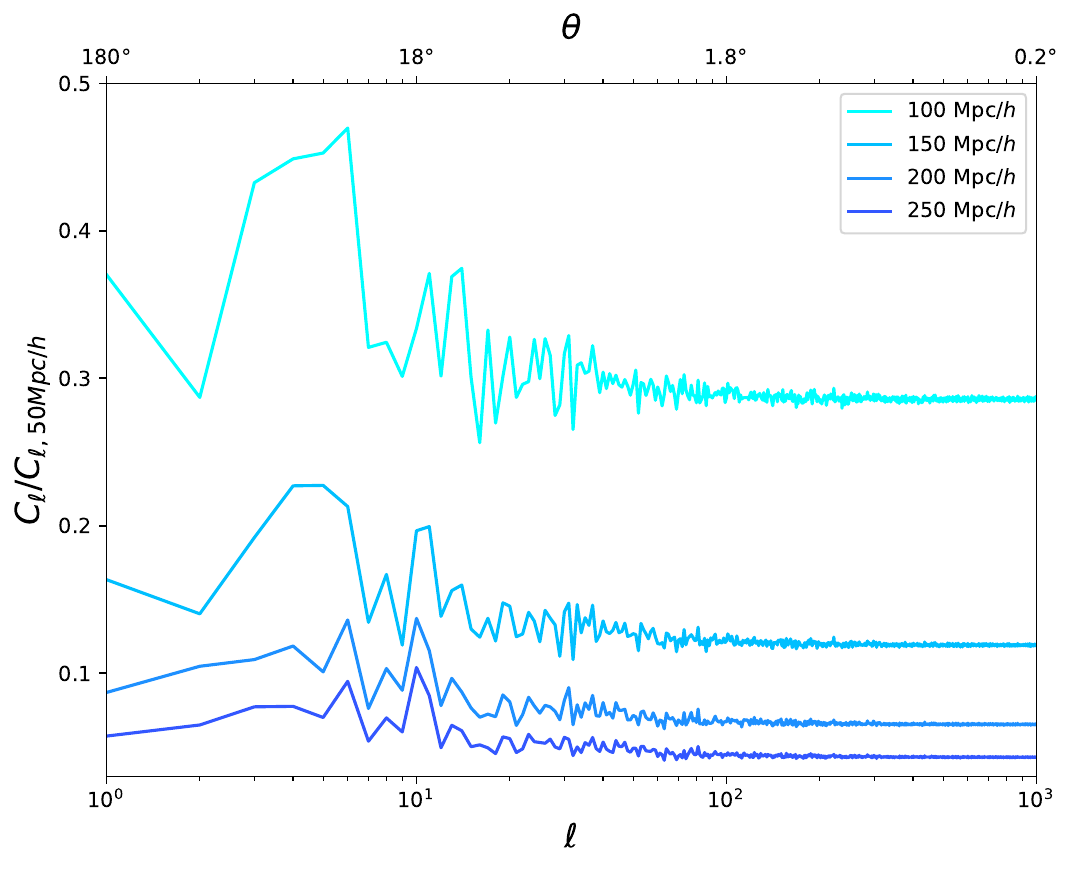}
	\quad
    \caption{The ratio of the angular power spectra of the overdensity $\delta_\mathrm{GW}$ in the SGWB from merging BBHs up to various comoving distances, relative to the angular power spectrum of the overdensity $\delta_\mathrm{GW}$ in the SGWB from merging BBHs up to $50$ Mpc/h, at $60$ Hz with $N_\mathrm{side}=512$ in the r2 model. The approximate scale of $\ell$ is indicated on the upper axis (rounded to one decimal place).}
    \label{fig:clsize}
\end{figure}

Fig. \ref{fig:clsize} presents the ratio of the angular power spectra of the overdensity $\delta_\mathrm{GW}$ in the SGWB from merging BBHs up to various comoving distances, relative to the angular power spectrum of the overdensity $\delta_\mathrm{GW}$ in the SGWB from merging BBHs up to $50$ Mpc/h, at $60$ Hz with $N_\mathrm{side}=512$ in the r2 model.
As sources at higher redshifts are included, the ratio decreases across $\ell = 1 - 1000$. For instance, the contribution from sources beyond $100$ Mpc/h to the total anisotropy is expected to be less than $\sim 30 \%-40 \%$. Meanwhile, the rate of decrease seems to slow. In comparison to the spectrum that incorporates sources up to $100$ Mpc/h ($\mathrm{C}_{\ell, 100 \mathrm{Mpc/h}}$), the spectrum that includes sources up to $150$ Mpc/h ($\mathrm{C}_{\ell, 150 \mathrm{Mpc/h}}$) shows a reduction of approximately $1.7-2.5$ times across the range $\ell = 1 - 1000$. However, when comparing the spectrum with sources up to $200$ Mpc/h ($\mathrm{C}_{\ell, 200 \mathrm{Mpc/h}}$) to that with sources up to $250$ Mpc/h ($\mathrm{C}_{\ell, 250 \mathrm{Mpc/h}}$), the latter exhibits a decrease of roughly $1.3-1.6$ times across the range $\ell = 1 - 1000$.  
Although a larger simulation box is necessary for a comprehensive exploration of the universe, our findings establish an upper limit for the anisotropy. 

\section{Summary and Discussion}
\label{sec:Summary and Discussion}
Anisotropic SGWB could serve as a probe of the spatial distribution of GW sources. We explore the anisotropic SGWB from local merging BBHs, NSBHs, and BNSs ($ d_c< 250 \mathrm{Mpc/h}$ ) by constructing seven all-sky mock lightcone GW event catalogs. The HEALPix skymaps of the overdensity $\delta_\mathrm{GW}$ in the anisotropic SGWB were generated using these mock catalogs. The HEALPix skymap of the r2 model for merging BBHs shown in Fig. \ref{fig:BBH GABE distribution} serves as an example to illustrate the anisotropic distribution of the GW sources. 

Based on the  HEALPix skymaps,  the angular power spectra in terms of  $\text{log}_{10} [\ell(\ell+1)\mathrm{C}_\ell/(2\pi)]$ are calculated to  quantify the anisotropy level of the SGWB from merging stellar-mass compact binaries. As shown in the upper panel of Fig. \ref{fig:cl}, the spectra for all three source types exhibit a typical shape at higher $\ell$ (e.g., $\ell > \sim 30 - 300$) across all catalogs, featuring an approximately linear increase with $\text{log}_{10} \ell$ : $\text{log}_{10} [\ell(\ell+1)\mathrm{C}_\ell/(2\pi)] \propto \beta(\text{log}_{10} \ell)$, with a characteristic slope of $\beta \sim 2$. The spectra for our seven catalogs exhibit considerable variations. For example, the spectra for merging BBHs in different catalogs vary by approximately a factor of $2$ to one order of magnitude for $\ell = 1$ to $\sim 80$, and by about $1$–$2$ orders of magnitude for $\ell = \sim 80$ to $1000$. The fluctuation in the distribution of the nearby (e.g., $< 50$ Mpc/h) sources should contribute to the variations of the spectra for merging BBHs, NSBHs and BNSs. Subtracting these nearby sources results in significant reductions of the variations of the spectra.   Shown  in the lower panel of Fig. \ref{fig:cl},  after the subtraction of the nearby sources, all spectra vary within a factor of $\sim 4$ for $\ell=1$ to $\sim 20$, and within a factor of $\sim 2$ for $\ell=\sim 20$ to $1000$. The spectra for the three source types then become closely aligned (within discrepancies of a factor of $\sim 2$ across $\ell = 1 - 1000$ for all catalogs).  This result indicates that the SGWB anisotropy is primarily influenced by the distribution of the hosts of GW sources and could be used as a probe of the LSS of the universe. 

We further examine the contribution of the radial distribution of GW sources to the anisotropy of the SGWB. The results for merging BBHs are shown in Fig.\ref{fig:clbins}. 
The random distance of the nearby sources makes the main contribution to the anisotropy of the SGWB (the red region in the upper panel of Fig.\ref{fig:clbins}). After the subtraction of the nearby sources, the radial distribution has an insignificant effect on the spectra (implied by the green region in the upper panel of Fig.\ref{fig:clbins}), and the anisotropy from the azimuthal distribution (approximately estimated using the 5-bin approximation) of sources varies within $\sim 3.5$ times across our seven mock catalogs (the blue region in the lower panel of Fig.\ref{fig:clbins}), for $\ell > \sim 10$. Therefore, the considerable spectral variations for $\ell>\sim 30$ shown in the upper panel of the Fig. \ref{fig:cl} could be attributed to the radial distribution of the nearby sources.  This random radial distribution of nearby sources might be the main Poisson noise, and contributes to the typical spectral shape as shown in \citet{2019PhRvD.100f3004C}. However, even with the extreme 1-bin approximation, namely the anisotropy approximately originating from the azimuthal distribution, the spectra continue to exhibit typical shapes for higher $\ell$ (e.g., $\ell > \sim 300 -600$). This indicate that the variation of spectra for the seven mock catalogs also include the Poisson noises from the azimuthal distribution.  We also find that including farther sources results in a rapid decrease in the anisotropy in Fig. \ref{fig:clsize}. However, the rates of decrease seem to diminish as sources at higher $z$ are included.

Besides the above results from the spectra for the merging BBHs shown in Figs.\ref{fig:clbins} and \ref{fig:clsize}, 
similar conclusions could also be obtained for both merging NSBHs and BNSs results. Furthermore, we investigate the effects of the observation frequency and the angular resolution of skymap on the spectra. We find that during the most sensitive observation frequency for LIGO, ranging from about $20$ to $100$ Hz, the spectra exhibit negligible variation. The spectra calculated from the HEALPix skymaps at different resolutions, specifically with $N_\mathrm{side} = 32, 64, 128, 256, 512$, are also consistent across the various configurations.

\section*{Acknowledgments}
We acknowledge valuable input from our anonymous referee. This work has been supported by the National Natural Science Foundation of China (Nos. 11988101, 11922303, 12033008 and 11673031), the K. C. Wong Education Foundation, and the National Key Research and Development Program of China (Grant Nos. 2020YFC2201400, SQ2021YFC220045-03). We acknowledge the use of the healpy and HEALPix packages in deriving the results presented in this paper.

\software{healpy \citep{Zonca2019}, HEALPix \citep{2005ApJ...622..759G}, GABE \citep{2019RAA....19..151J}, COSMIC v3.3.0 \citep{2020ApJ...898...71B, katie_breivik_2020_3905335}, NumPy \citep{2020NumPy-Array}, Matplotlib \citep{Hunter:2007}, Astropy \citep{astropy:2013, astropy:2018, astropy:2022}, SciPy \citep{2020SciPy-NMeth}, Numba \citep{2015llvm.confE...1L}, Jupyter \citep{2016ppap.book...87K}, Python (https://www.python.org)}

\bibliography{main}{}

\begin{thebibliography}{}
\expandafter\ifx\csname natexlab\endcsname\relax\def\natexlab#1{#1}\fi
\providecommand{\url}[1]{\href{#1}{#1}}
\providecommand{\dodoi}[1]{doi:~\href{http://doi.org/#1}{\nolinkurl{#1}}}
\providecommand{\doeprint}[1]{\href{http://ascl.net/#1}{\nolinkurl{http://ascl.net/#1}}}
\providecommand{\doarXiv}[1]{\href{https://arxiv.org/abs/#1}{\nolinkurl{https://arxiv.org/abs/#1}}}

\bibitem[{{Abbott} {et~al.}(2021){Abbott}, {Abbott}, {Abraham}, {Acernese},
  {Ackley}, {Adams}, {Adams}, {Adhikari}, {Adya}, {Affeldt}, {Agarwal},
  {Agathos}, {Agatsuma}, {Aggarwal}, {Aguiar}, {Aiello}, {Ain}, {Akutsu},
  {Aleman}, {Allen}, {Allocca}, {Altin}, {Amato}, {Anand}, {Ananyeva},
  {Anderson}, {Anderson}, {Ando}, {Angelova}, {Ansoldi}, {Antelis}, {Antier},
  {Appert}, {Arai}, {Arai}, {Arai}, {Araki}, {Araya}, {Araya}, {Areeda},
  {Ar{\`e}ne}, {Aritomi}, {Arnaud}, {Aronson}, {Asada}, {Asali}, {Ashton},
  {Aso}, {Aston}, {Astone}, {Aubin}, {Aufmuth}, {Aultoneal}, {Austin}, {Babak},
  {Badaracco}, {Bader}, {Bae}, {Bae}, {Baer}, {Bagnasco}, {Bai}, {Baiotti},
  {Baird}, {Bajpai}, {Ball}, {Ballardin}, {Ballmer}, {Bals}, {Balsamo},
  {Baltus}, {Banagiri}, {Bankar}, {Bankar}, {Barayoga}, {Barbieri}, {Barish},
  {Barker}, {Barneo}, {Barnum}, {Barone}, {Barr}, {Barsotti}, {Barsuglia},
  {Barta}, {Bartlett}, {Barton}, {Bartos}, {Bassiri}, {Basti}, {Bawaj},
  {Bayley}, {Baylor}, {Bazzan}, {B{\'e}csy}, {Bedakihale}, {Bejger},
  {Belahcene}, {Benedetto}, {Beniwal}, {Benjamin}, {Bennett}, {Bentley},
  {Benyaala}, {Bergamin}, {Berger}, {Bernuzzi}, {Bersanetti}, {Bertolini},
  {Betzwieser}, {Bhandare}, {Bhandari}, {Bhattacharjee}, {Bhaumik}, {Bidler},
  {Bilenko}, {Billingsley}, {Birney}, {Birnholtz}, {Biscans}, {Bischi},
  {Biscoveanu}, {Bisht}, {Biswas}, {Bitossi}, {Bizouard}, {Blackburn},
  {Blackman}, {Blair}, {Blair}, {Blair}, {Bobba}, {Bode}, {Boer}, {Bogaert},
  {Boldrini}, {Bondu}, {Bonilla}, {Bonnand}, {Booker}, {Boom}, {Bork},
  {Boschi}, {Bose}, {Bose}, {Bossilkov}, {Boudart}, {Bouffanais}, {Bozzi},
  {Bradaschia}, {Brady}, {Bramley}, {Branch}, {Branchesi}, {Brau}, {Breschi},
  {Briant}, {Briggs}, {Brillet}, {Brinkmann}, {Brockill}, {Brooks}, {Brooks},
  {Brown}, {Brunett}, {Bruno}, {Bruntz}, {Bryant}, {Buikema}, {Bulik},
  {Bulten}, {Buonanno}, {Buscicchio}, {Buskulic}, {Byer}, {Cadonati}, {Caesar},
  {Cagnoli}, {Cahillane}, {Cain}, {Bustillo}, {Callaghan}, {Callister},
  {Calloni}, {Camp}, {Canepa}, {Cannavacciuolo}, {Cannon}, {Cao}, {Cao}, {Cao},
  {Capocasa}, {Capote}, {Carapella}, {Carbognani}, {Carlin}, {Carney},
  {Carpinelli}, {Carullo}, {Carver}, {Diaz}, {Casentini}, {Castaldi},
  {Caudill}, {Cavagli{\`a}}, {Cavalier}, {Cavalieri}, {Cella},
  {Cerd{\'a}-Dur{\'a}n}, {Cesarini}, {Chaibi}, {Chakravarti}, {Champion},
  {Chan}, {Chan}, {Chan}, {Chan}, {Chandra}, {Chanial}, {Chao}, {Charlton},
  {Chase}, {Chassande-Mottin}, {Chatterjee}, {Chaturvedi}, {Chen}, {Chen},
  {Chen}, {Chen}, {Chen}, {Chen}, {Chen}, {Chen}, {Chen}, {Cheng}, {Cheong},
  {Cheung}, {Chia}, {Chiadini}, {Chiang}, {Chierici}, {Chincarini}, {Chiofalo},
  {Chiummo}, {Cho}, {Cho}, {Choate}, {Choudhary}, {Choudhary}, {Christensen},
  {Chu}, {Chu}, {Chu}, {Chua}, {Chung}, {Ciani}, {Ciecielag}, {Cie{\'s}lar},
  {Cifaldi}, {Ciobanu}, {Ciolfi}, {Cipriano}, {Cirone}, {Clara}, {Clark},
  {Clark}, {Clarke}, {Clearwater}, {Clesse}, {Cleva}, {Coccia}, {Cohadon},
  {Cohen}, {Cohen}, {Colleoni}, {Collette}, {Colpi}, {Compton}, {Constancio},
  {Conti}, {Cooper}, {Corban}, {Corbitt}, {Cordero-Carri{\'o}n}, {Corezzi},
  {Corley}, {Cornish}, {Corre}, {Corsi}, {Cortese}, {Costa}, {Cotesta},
  {Coughlin}, {Coughlin}, {Coulon}, {Countryman}, {Cousins}, {Couvares},
  {Covas}, {Coward}, {Cowart}, {Coyne}, {Coyne}, {Creighton}, {Creighton},
  {Criswell}, {Croquette}, {Crowder}, {Cudell}, {Cullen}, {Cumming},
  {Cummings}, {Cuoco}, {Cury{\l}o}, {Canton}, {D{\'a}lya}, {Dana},
  {Daneshgaranbajastani}, {D'Angelo}, {Danilishin}, {D'Antonio}, {Danzmann},
  {Darsow-Fromm}, {Dasgupta}, {Datrier}, {Dattilo}, {Dave}, {Davier}, {Davies},
  {Davis}, {Daw}, {Dean}, {Deenadayalan}, {Degallaix}, {de Laurentis},
  {Del{\'e}glise}, {Del Favero}, {de Lillo}, {de Lillo}, {Del Pozzo},
  {Demarchi}, {de Matteis}, {D'Emilio}, {Demos}, {Dent}, {Depasse}, {de
  Pietri}, {De Rosa}, {de Rossi}, {Desalvo}, {de Simone}, {Dhurandhar},
  {D{\'\i}az}, {Diaz-Ortiz}, {Didio}, {Dietrich}, {di Fiore}, {di Fronzo}, {di
  Giorgio}, {di Giovanni}, {di Girolamo}, {di Lieto}, {Ding}, {di Pace}, {di
  Palma}, {di Renzo}, {Divakarla}, {Dmitriev}, {Doctor}, {D'Onofrio},
  {Donovan}, {Dooley}, {Doravari}, {Dorrington}, {Drago}, {Driggers}, {Drori},
  {Du}, {Ducoin}, {Dupej}, {Durante}, {D'Urso}, {Duverne}, {Dvorkin}, {Dwyer},
  {Easter}, {Ebersold}, {Eddolls}, {Edelman}, {Edo}, {Edy}, {Effler}, {Eguchi},
  {Eichholz}, {Eikenberry}, {Eisenmann}, {Eisenstein}, {Ejlli}, {Enomoto},
  {Errico}, {Essick}, {Estell{\'e}s}, {Estevez}, {Etienne}, {Etzel}, {Evans},
  {Evans}, {Ewing}, {Fafone}, {Fair}, {Fairhurst}, {Fan}, {Farah}, {Farinon},
  {Farr}, {Farr}, {Farrow}, {Fauchon-Jones}, {Favata}, {Fays}, {Fazio},
  {Feicht}, {Fejer}, {Feng}, {Fenyvesi}, {Ferguson}, {Fernandez-Galiana},
  {Ferrante}, {Ferreira}, {Fidecaro}, {Figura}, {Fiori}, {Fishbach}, {Fisher},
  {Fishner}, {Fittipaldi}, {Fiumara}, {Flaminio}, {Floden}, {Flynn}, {Fong},
  {Font}, {Fornal}, {Forsyth}, {Franke}, {Frasca}, {Frasconi}, {Frederick},
  {Frei}, {Freise}, {Frey}, {Fritschel}, {Frolov}, {Fronz{\'e}}, {Fujii},
  {Fujikawa}, {Fukunaga}, {Fukushima}, {Fulda}, {Fyffe}, {Gabbard}, {Gadre},
  {Gaebel}, {Gair}, {Gais}, {Galaudage}, {Gamba}, {Ganapathy}, {Ganguly},
  {Gao}, {Gaonkar}, {Garaventa}, {Garc{\'\i}a-N{\'u}{\~n}ez},
  {Garc{\'\i}a-Quir{\'o}s}, {Garufi}, {Gateley}, {Gaudio}, {Gayathri}, {Ge},
  {Gemme}, {Gennai}, {George}, {Gergely}, {Gewecke}, {Ghonge}, {Ghosh},
  {Ghosh}, {Ghosh}, {Ghosh}, {Ghosh}, {Giacomazzo}, {Giacoppo}, {Giaime},
  {Giardina}, {Gibson}, {Gier}, {Giesler}, {Giri}, {Gissi}, {Glanzer},
  {Gleckl}, {Godwin}, {Goetz}, {Goetz}, {Gohlke}, {Goncharov}, {Gonz{\'a}lez},
  {Gopakumar}, {Gosselin}, {Gouaty}, {Grace}, {Grado}, {Granata}, {Granata},
  {Grant}, {Gras}, {Grassia}, {Gray}, {Gray}, {Greco}, {Green}, {Green},
  {Gretarsson}, {Gretarsson}, {Griffith}, {Griffiths}, {Griggs}, {Grignani},
  {Grimaldi}, {Grimes}, {Grimm}, {Grote}, {Grunewald}, {Gruning}, {Guerrero},
  {Guidi}, {Guimaraes}, {Guix{\'e}}, {Gulati}, {Guo}, {Guo}, {Gupta}, {Gupta},
  {Gupta}, {Gustafson}, {Gustafson}, {Guzman}, {Ha}, {Haegel}, {Hagiwara},
  {Haino}, {Halim}, {Hall}, {Hamilton}, {Hammond}, {Han}, {Haney}, {Hanks},
  {Hanna}, {Hannam}, {Hannuksela}, {Hansen}, {Hansen}, {Hanson}, {Harder},
  {Hardwick}, {Haris}, {Harms}, {Harry}, {Harry}, {Hartwig}, {Hasegawa},
  {Haskell}, {Hasskew}, {Haster}, {Hattori}, {Haughian}, {Hayakawa}, {Hayama},
  {Hayes}, {Healy}, {Heidmann}, {Heintze}, {Heinze}, {Heinzel}, {Heitmann},
  {Hellman}, {Hello}, {Helmling-Cornell}, {Hemming}, {Hendry}, {Heng},
  {Hennes}, {Hennig}, {Hennig}, {Vivanco}, {Heurs}, {Hild}, {Hill}, {Himemoto},
  {Hines}, {Hiranuma}, {Hirata}, {Hirose}, {Hochheim}, {Hofman}, {Hohmann},
  {Holgado}, {Holland}, {Hollows}, {Holmes}, {Holt}, {Holz}, {Hong}, {Hopkins},
  {Hough}, {Howell}, {Hoy}, {Hoyland}, {Hreibi}, {Hsieh}, {Hsu}, {Huang},
  {Huang}, {Huang}, {Huang}, {Huang}, {Huang}, {H{\"u}bner}, {Huddart},
  {Huerta}, {Hughey}, {Hui}, {Hui}, {Husa}, {Huttner}, {Huxford}, {Huynh-Dinh},
  {Ide}, {Idzkowski}, {Iess}, {Ikenoue}, {Imam}, {Inayoshi}, {Inchauspe},
  {Ingram}, {Inoue}, {Intini}, {Ioka}, {Isi}, {Isleif}, {Ito}, {Itoh}, {Iyer},
  {Izumi}, {Jaberianhamedan}, {Jacqmin}, {Jadhav}, {Jadhav}, {James}, {Jan},
  {Jani}, {Janssens}, {Janthalur}, {Jaranowski}, {Jariwala}, {Jaume},
  {Jenkins}, {Jeon}, {Jeunon}, {Jia}, {Jiang}, {Jin}, {Johns}, {Jones},
  {Jones}, {Jones}, {Jones}, {Jones}, {Jonker}, {Ju}, {Jung}, {Jung}, {Junker},
  {Kaihotsu}, {Kajita}, {Kakizaki}, {Kalaghatgi}, {Kalogera}, {Kamai},
  {Kamiizumi}, {Kanda}, {Kandhasamy}, {Kang}, {Kanner}, {Kao}, {Kapadia},
  {Kapasi}, {Karathanasis}, {Karki}, {Kashyap}, {Kasprzack}, {Kastaun},
  {Katsanevas}, {Katsavounidis}, {Katzman}, {Kaur}, {Kawabe}, {Kawaguchi},
  {Kawai}, {Kawasaki}, {K{\'e}f{\'e}lian}, {Keitel}, {Key}, {Khadka},
  {Khalili}, {Khan}, {Khan}, {Khazanov}, {Khetan}, {Khursheed}, {Kijbunchoo},
  {Kim}, {Kim}, {Kim}, {Kim}, {Kim}, {Kim}, {Kimball}, {Kimura}, {King},
  {Kinley-Hanlon}, {Kirchhoff}, {Kissel}, {Kita}, {Kitazawa}, {Kleybolte},
  {Klimenko}, {Knee}, {Knowles}, {Knyazev}, {Koch}, {Koekoek}, {Kojima},
  {Kokeyama}, {Koley}, {Kolitsidou}, {Kolstein}, {Komori}, {Kondrashov},
  {Kong}, {Kontos}, {Koper}, {Korobko}, {Kotake}, {Kovalam}, {Kozak},
  {Kozakai}, {Kozu}, {Kringel}, {Krishnendu}, {Kr{\'o}lak}, {Kuehn}, {Kuei},
  {Kumar}, {Kumar}, {Kumar}, {Kumar}, {Kume}, {Kuns}, {Kuo}, {Kuo}, {Kuromiya},
  {Kuroyanagi}, {Kusayanagi}, {Kwak}, {Kwang}, {Laghi}, {Lalande}, {Lam},
  {Lamberts}, {Landry}, {Lane}, {Lang}, {Lange}, {Lantz}, {La Rosa},
  {Lartaux-Vollard}, {Lasky}, {Laxen}, {Lazzarini}, {Lazzaro}, {Leaci},
  {Leavey}, {Lecoeuche}, {Lee}, {Lee}, {Lee}, {Lee}, {Lee}, {Lee}, {Lehmann},
  {Lema{\^\i}tre}, {Leon}, {Leonardi}, {Leroy}, {Letendre}, {Levin}, {Leviton},
  {Li}, {Li}, {Li}, {Li}, {Li}, {Li}, {Lin}, {Lin}, {Lin}, {Lin}, {Lin},
  {Linde}, {Linker}, {Linley}, {Littenberg}, {Liu}, {Liu}, {Liu}, {Liu},
  {Llorens-Monteagudo}, {Lo}, {Lockwood}, {Lollie}, {London}, {Longo}, {Lopez},
  {Lorenzini}, {Loriette}, {Lormand}, {Losurdo}, {Lough}, {Lousto}, {Lovelace},
  {L{\"u}ck}, {Lumaca}, {Lundgren}, {Luo}, {Macas}, {Macinnis}, {MacLeod},
  {MacMillan}, {Macquet}, {Hernandez}, {Maga{\~n}a-Sandoval}, {Magazz{\`u}},
  {Magee}, {Maggiore}, {Majorana}, {Maksimovic}, {Maliakal}, {Malik}, {Man},
  {Mandic}, {Mangano}, {Mango}, {Mansell}, {Manske}, {Mantovani}, {Mapelli},
  {Marchesoni}, {Marchio}, {Marion}, {Mark}, {M{\'a}rka}, {M{\'a}rka},
  {Markakis}, {Markosyan}, {Markowitz}, {Maros}, {Marquina}, {Marsat},
  {Martelli}, {Martin}, {Martin}, {Martinez}, {Martinez}, {Martinovic},
  {Martynov}, {Marx}, {Masalehdan}, {Mason}, {Massera}, {Masserot},
  {Massinger}, {Masso-Reid}, {Mastrogiovanni}, {Matas}, {Mateu-Lucena},
  {Matichard}, {Matiushechkina}, {Mavalvala}, {McCann}, {McCarthy},
  {McClelland}, {McClincy}, {McCormick}, {McCuller}, {McGhee}, {McGuire},
  {McIsaac}, {McIver}, {McManus}, {McRae}, {McWilliams}, {Meacher}, {Mehmet},
  {Mehta}, {Melatos}, {Melchor}, {Mendell}, {Menendez-Vazquez}, {Menoni},
  {Mercer}, {Mereni}, {Merfeld}, {Merilh}, {Merritt}, {Merzougui}, {Meshkov},
  {Messenger}, {Messick}, {Meyers}, {Meylahn}, {Mhaske}, {Miani}, {Miao},
  {Michaloliakos}, {Michel}, {Michimura}, {Middleton}, {Milano}, {Miller},
  {Millhouse}, {Mills}, {Milotti}, {Milovich-Goff}, {Minazzoli}, {Minenkov},
  {Mio}, {Mir}, {Mishkin}, {Mishra}, {Mishra}, {Mistry}, {Mitra}, {Mitrofanov},
  {Mitselmakher}, {Mittleman}, {Miyakawa}, {Miyamoto}, {Miyazaki}, {Miyo},
  {Miyoki}, {Mo}, {Mogushi}, {Mohapatra}, {Mohite}, {Molina}, {Molina-Ruiz},
  {Mondin}, {Montani}, {Moore}, {Moraru}, {Morawski}, {More}, {Moreno},
  {Moreno}, {Mori}, {Morisaki}, {Moriwaki}, {Mours}, {Mow-Lowry}, {Mozzon},
  {Muciaccia}, {Mukherjee}, {Mukherjee}, {Mukherjee}, {Mukherjee}, {Mukund},
  {Mullavey}, {Munch}, {Mu{\~n}iz}, {Murray}, {Musenich}, {Nadji}, {Nagano},
  {Nagano}, {Nagar}, {Nakamura}, {Nakano}, {Nakano}, {Nakashima}, {Nakayama},
  {Nardecchia}, {Narikawa}, {Naticchioni}, {Nayak}, {Nayak}, {Negishi}, {Neil},
  {Neilson}, {Nelemans}, {Nelson}, {Nery}, {Neunzert}, {Ng}, {Ng}, {Nguyen},
  {Nguyen}, {Nguyen}, {Quynh}, {Ni}, {Nichols}, {Nishizawa}, {Nissanke},
  {Nocera}, {Noh}, {Norman}, {North}, {Nozaki}, {Nuttall}, {Oberling},
  {O'Brien}, {Obuchi}, {O'Dell}, {Ogaki}, {Oganesyan}, {Oh}, {Oh}, {Oh},
  {Ohashi}, {Ohishi}, {Ohkawa}, {Ohme}, {Ohta}, {Okada}, {Okutani}, {Okutomi},
  {Olivetto}, {Oohara}, {Ooi}, {Oram}, {O'Reilly}, {Ormiston}, {Ormsby},
  {Ortega}, {O'Shaughnessy}, {O'Shea}, {Oshino}, {Ossokine}, {Osthelder},
  {Otabe}, {Ottaway}, {Overmier}, {Pace}, {Pagano}, {Page}, {Pagliaroli},
  {Pai}, {Pai}, {Palamos}, {Palashov}, {Palomba}, {Pan}, {Panda}, {Pang},
  {Pang}, {Pankow}, {Pannarale}, {Pant}, {Paoletti}, {Paoli}, {Paolone},
  {Parisi}, {Park}, {Parker}, {Pascucci}, {Pasqualetti}, {Passaquieti},
  {Passuello}, {Patel}, {Patricelli}, {Payne}, {Pechsiri}, {Pedraza},
  {Pegoraro}, {Pele}, {Arellano}, {Penn}, {Perego}, {Pereira}, {Pereira},
  {Perez}, {P{\'e}rigois}, {Perreca}, {Perri{\`e}s}, {Petermann}, {Petterson},
  {Pfeiffer}, {Pham}, {Phukon}, {Piccinni}, {Pichot}, {Piendibene},
  {Piergiovanni}, {Pierini}, {Pierro}, {Pillant}, {Pilo}, {Pinard}, {Pinto},
  {Piotrzkowski}, {Piotrzkowski}, {Pirello}, {Pitkin}, {Placidi}, {Plastino},
  {Pluchar}, {Poggiani}, {Polini}, {Pong}, {Ponrathnam}, {Popolizio}, {Porter},
  {Powell}, {Pracchia}, {Pradier}, {Prajapati}, {Prasai}, {Prasanna},
  {Pratten}, {Prestegard}, {Principe}, {Prodi}, {Prokhorov}, {Prosposito},
  {Prudenzi}, {Puecher}, {Punturo}, {Puosi}, {Puppo}, {P{\"u}rrer}, {Qi},
  {Quetschke}, {Quinonez}, {Quitzow-James}, {Raab}, {Raaijmakers}, {Radkins},
  {Radulesco}, {Raffai}, {Rail}, {Raja}, {Rajan}, {Ramirez}, {Ramirez},
  {Ramos-Buades}, {Rana}, {Rapagnani}, {Rapol}, {Ratto}, {Raymond}, {Raza},
  {Razzano}, {Read}, {Rees}, {Regimbau}, {Rei}, {Reid}, {Reitze}, {Relton},
  {Rettegno}, {Ricci}, {Richardson}, {Richardson}, {Richardson}, {Ricker},
  {Riemenschneider}, {Riles}, {Rizzo}, {Robertson}, {Robie}, {Robinet},
  {Rocchi}, {Rocha}, {Rodriguez}, {Rodriguez-Soto}, {Rolland}, {Rollins},
  {Roma}, {Romanelli}, {Romano}, {Romano}, {Romel}, {Romero}, {Romero-Shaw},
  {Romie}, {Rose}, {Rosi{\'n}ska}, {Rosofsky}, {Ross}, {Rowan}, {Rowlinson},
  {Roy}, {Roy}, {Rozza}, {Ruggi}, {Ryan}, {Sachdev}, {Sadecki}, {Sadiq},
  {Sago}, {Saito}, {Saito}, {Sakai}, {Sakai}, {Sakellariadou}, {Sakuno},
  {Salafia}, {Salconi}, {Saleem}, {Salemi}, {Samajdar}, {Sanchez}, {Sanchez},
  {Sanchez}, {Sanchis-Gual}, {Sanders}, {Sanuy}, {Saravanan}, {Sarin},
  {Sassolas}, {Satari}, {Sato}, {Sato}, {Sauter}, {Savage}, {Savant}, {Sawada},
  {Sawant}, {Sawant}, {Sayah}, {Schaetzl}, {Scheel}, {Scheuer},
  {Schindler-Tyka}, {Schmidt}, {Schnabel}, {Schneewind}, {Schofield},
  {Sch{\"o}nbeck}, {Schulte}, {Schutz}, {Schwartz}, {Scott}, {Scott},
  {Seglar-Arroyo}, {Seidel}, {Sekiguchi}, {Sekiguchi}, {Sellers}, {Sengupta},
  {Sennett}, {Sentenac}, {Seo}, {Sequino}, {Sergeev}, {Setyawati}, {Shaffer},
  {Shahriar}, {Shams}, {Shao}, {Sharifi}, {Sharma}, {Sharma}, {Shawhan},
  {Shcheblanov}, {Shen}, {Shibagaki}, {Shikauchi}, {Shimizu}, {Shimoda},
  {Shimode}, {Shink}, {Shinkai}, {Shishido}, {Shoda}, {Shoemaker}, {Shoemaker},
  {Shukla}, {Shyamsundar}, {Sieniawska}, {Sigg}, {Singer}, {Singh}, {Singh},
  {Singha}, {Sintes}, {Sipala}, {Skliris}, {Slagmolen}, {Slaven-Blair},
  {Smetana}, {Smith}, {Smith}, {Somala}, {Somiya}, {Son}, {Soni}, {Soni},
  {Sorazu}, {Sordini}, {Sorrentino}, {Sorrentino}, {Sotani}, {Soulard},
  {Souradeep}, {Sowell}, {Spagnuolo}, {Spencer}, {Spera}, {Srivastava},
  {Srivastava}, {Staats}, {Stachie}, {Steer}, {Steinlechner}, {Steinlechner},
  {Stops}, {Stover}, {Strain}, {Strang}, {Stratta}, {Strunk}, {Sturani},
  {Stuver}, {S{\"u}dbeck}, {Sudhagar}, {Sudhir}, {Sugimoto}, {Suh},
  {Summerscales}, {Sun}, {Sun}, {Sunil}, {Sur}, {Suresh}, {Sutton}, {Suzuki},
  {Suzuki}, {Swinkels}, {Szczepa{\'n}czyk}, {Szewczyk}, {Tacca}, {Tagoshi},
  {Tait}, {Takahashi}, {Takahashi}, {Takamori}, {Takano}, {Takeda}, {Takeda},
  {Talbot}, {Tanaka}, {Tanaka}, {Tanaka}, {Tanaka}, {Tanaka}, {Tanasijczuk},
  {Tanioka}, {Tanner}, {Tao}, {Tapia}, {Martin}, {Martin}, {Tasson}, {Telada},
  {Tenorio}, {Terkowski}, {Test}, {Thirugnanasambandam}, {Thomas}, {Thomas},
  {Thompson}, {Thondapu}, {Thorne}, {Thrane}, {Tiwari}, {Tiwari}, {Tiwari},
  {Toland}, {Tolley}, {Tomaru}, {Tomigami}, {Tomura}, {Tonelli},
  {Torres-Forn{\'e}}, {Torrie}, {E Melo}, {T{\"o}yr{\"a}}, {Trapananti},
  {Travasso}, {Traylor}, {Tringali}, {Tripathee}, {Troiano}, {Trovato},
  {Trozzo}, {Trudeau}, {Tsai}, {Tsai}, {Tsang}, {Tsang}, {Tsao}, {Tse}, {Tso},
  {Tsubono}, {Tsuchida}, {Tsukada}, {Tsuna}, {Tsutsui}, {Tsuzuki}, {Turconi},
  {Tuyenbayev}, {Ubhi}, {Uchikata}, {Uchiyama}, {Udall}, {Ueda}, {Uehara},
  {Ueno}, {Ueshima}, {Ugolini}, {Unnikrishnan}, {Uraguchi}, {Urban}, {Ushiba},
  {Usman}, {Utina}, {Vahlbruch}, {Vajente}, {Vajpeyi}, {Valdes}, {Valentini},
  {Valsan}, {van Bakel}, {van Beuzekom}, {van den Brand}, {van den Broeck},
  {van Remortel}, {Vander-Hyde}, {van der Schaaf}, {van Heijningen}, {van
  Putten}, {Vardaro}, {Vargas}, {Varma}, {Vas{\'u}th}, {Vecchio}, {Vedovato},
  {Veitch}, {Veitch}, {Venkateswara}, {Venneberg}, {Venugopalan}, {Verkindt},
  {Verma}, {Veske}, {Vetrano}, {Vicer{\'e}}, {Viets}, {Villa-Ortega}, {Vinet},
  {Vitale}, {Vo}, {Vocca}, {von Reis}, {von Wrangel}, {Vorvick}, {Vyatchanin},
  {Wade}, {Wade}, {Wagner}, {Walet}, {Walker}, {Wallace}, {Wallace}, {Walsh},
  {Wang}, {Wang}, {Wang}, {Ward}, {Warner}, {Was}, {Washimi}, {Washington},
  {Watchi}, {Weaver}, {Wei}, {Weinert}, {Weinstein}, {Weiss}, {Weller},
  {Wellmann}, {Wen}, {We{\ss}els}, {Westhouse}, {Wette}, {Whelan}, {White},
  {Whiting}, {Whittle}, {Wilken}, {Williams}, {Williams}, {Williamson},
  {Willis}, {Willke}, {Wilson}, {Winkler}, {Wipf}, {Wlodarczyk}, {Woan},
  {Woehler}, {Wofford}, {Wong}, {Wu}, {Wu}, {Wu}, {Wu}, {Wysocki}, {Xiao},
  {Xu}, {Yamada}, {Yamamoto}, {Yamamoto}, {Yamamoto}, {Yamamoto}, {Yamashita},
  {Yamazaki}, {Yang}, {Yang}, {Yang}, {Yang}, {Yang}, {Yap}, {Yeeles},
  {Yelikar}, {Ying}, {Yokogawa}, {Yokoyama}, {Yokozawa}, {Yoon}, {Yoshioka},
  {Yu}, {Yu}, {Yuzurihara}, {Zadro{\.z}ny}, {Zanolin}, {Zeidler}, {Zelenova},
  {Zendri}, {Zevin}, {Zhan}, {Zhang}, {Zhang}, {Zhang}, {Zhang}, {Zhang},
  {Zhao}, {Zhao}, {Zhao}, {Zhao}, {Zhou}, {Zhu}, {Zhu}, {Zucker}, {Zweizig},
  {Ligo Scientific Collaboration}, {VIRGO Collaboration}, \& {Kagra
  Collaboration}}]{2021PhRvD.104b2004A}
{Abbott}, R., {Abbott}, T.~D., {Abraham}, S., {et~al.} 2021, \prd, 104, 022004,
  \dodoi{10.1103/PhysRevD.104.022004}

\bibitem[{{Abbott} {et~al.}(2023{\natexlab{a}}){Abbott}, {Abbott}, {Acernese},
  {Ackley}, {Adams}, {Adhikari}, {Adhikari}, {Adya}, {Affeldt}, {Agarwal},
  {Agathos}, {Agatsuma}, {Aggarwal}, {Aguiar}, {Aiello}, {Ain}, {Ajith},
  {Akcay}, {Akutsu}, {Albanesi}, {Allocca}, {Altin}, {Amato}, {Anand}, {Anand},
  {Ananyeva}, {Anderson}, {Anderson}, {Ando}, {Andrade}, {Andres},
  {Andri{\'c}}, {Angelova}, {Ansoldi}, {Antelis}, {Antier}, {Appert}, {Arai},
  {Arai}, {Arai}, {Araki}, {Araya}, {Araya}, {Areeda}, {Ar{\`e}ne}, {Aritomi},
  {Arnaud}, {Arogeti}, {Aronson}, {Arun}, {Asada}, {Asali}, {Ashton}, {Aso},
  {Assiduo}, {Aston}, {Astone}, {Aubin}, {Austin}, {Babak}, {Badaracco},
  {Bader}, {Badger}, {Bae}, {Bae}, {Baer}, {Bagnasco}, {Bai}, {Baiotti},
  {Baird}, {Bajpai}, {Ball}, {Ballardin}, {Ballmer}, {Balsamo}, {Baltus},
  {Banagiri}, {Bankar}, {Barayoga}, {Barbieri}, {Barish}, {Barker}, {Barneo},
  {Barone}, {Barr}, {Barsotti}, {Barsuglia}, {Barta}, {Bartlett}, {Barton},
  {Bartos}, {Bassiri}, {Basti}, {Bawaj}, {Bayley}, {Baylor}, {Bazzan},
  {B{\'e}csy}, {Bedakihale}, {Bejger}, {Belahcene}, {Benedetto}, {Beniwal},
  {Bennett}, {Bentley}, {Benyaala}, {Bergamin}, {Berger}, {Bernuzzi}, {Berry},
  {Bersanetti}, {Bertolini}, {Betzwieser}, {Beveridge}, {Bhandare}, {Bhardwaj},
  {Bhattacharjee}, {Bhaumik}, {Bilenko}, {Billingsley}, {Bini}, {Birney},
  {Birnholtz}, {Biscans}, {Bischi}, {Biscoveanu}, {Bisht}, {Biswas}, {Bitossi},
  {Bizouard}, {Blackburn}, {Blair}, {Blair}, {Blair}, {Bobba}, {Bode}, {Boer},
  {Bogaert}, {Boldrini}, {Bonavena}, {Bondu}, {Bonilla}, {Bonnand}, {Booker},
  {Boom}, {Bork}, {Boschi}, {Bose}, {Bose}, {Bossilkov}, {Boudart},
  {Bouffanais}, {Bozzi}, {Bradaschia}, {Brady}, {Bramley}, {Branch},
  {Branchesi}, {Brandt}, {Brau}, {Breschi}, {Briant}, {Briggs}, {Brillet},
  {Brinkmann}, {Brockill}, {Brooks}, {Brooks}, {Brown}, {Brunett}, {Bruno},
  {Bruntz}, {Bryant}, {Bulik}, {Bulten}, {Buonanno}, {Buscicchio}, {Buskulic},
  {Buy}, {Byer}, {Davies}, {Cadonati}, {Cagnoli}, {Cahillane}, {Bustillo},
  {Callaghan}, {Callister}, {Calloni}, {Cameron}, {Camp}, {Canepa},
  {Canevarolo}, {Cannavacciuolo}, {Cannon}, {Cao}, {Cao}, {Capocasa}, {Capote},
  {Carapella}, \& {Carbognani}}]{2023PhRvX..13d1039A}
{Abbott}, R., {Abbott}, T.~D., {Acernese}, F., {et~al.} 2023{\natexlab{a}},
  Physical Review X, 13, 041039, \dodoi{10.1103/PhysRevX.13.041039}

\bibitem[{{Abbott} {et~al.}(2023{\natexlab{b}}){Abbott}, {Abbott}, {Acernese},
  {Ackley}, {Adams}, {Adhikari}, {Adhikari}, {Adya}, {Affeldt}, {Agarwal},
  {Agathos}, {Agatsuma}, {Aggarwal}, {Aguiar}, {Aiello}, {Ain}, {Ajith},
  {Akutsu}, {de Alarc{\'o}n}, {Akcay}, {Albanesi}, {Allocca}, {Altin}, {Amato},
  {Anand}, {Anand}, {Ananyeva}, {Anderson}, {Anderson}, {Ando}, {Andrade},
  {Andres}, {Andri{\'c}}, {Angelova}, {Ansoldi}, {Antelis}, {Antier},
  {Antonini}, {Appert}, {Arai}, {Arai}, {Arai}, {Araki}, {Araya}, {Araya},
  {Areeda}, {Ar{\`e}ne}, {Aritomi}, {Arnaud}, {Arogeti}, {Aronson}, {Arun},
  {Asada}, {Asali}, {Ashton}, {Aso}, {Assiduo}, {Aston}, {Astone}, {Aubin},
  {Austin}, {Babak}, {Badaracco}, {Bader}, {Badger}, {Bae}, {Bae}, {Baer},
  {Bagnasco}, {Bai}, {Baiotti}, {Baird}, {Bajpai}, {Ball}, {Ballardin},
  {Ballmer}, {Balsamo}, {Baltus}, {Banagiri}, {Bankar}, {Barayoga}, {Barbieri},
  {Barish}, {Barker}, {Barneo}, {Barone}, {Barr}, {Barsotti}, {Barsuglia},
  {Barta}, {Bartlett}, {Barton}, {Bartos}, {Bassiri}, {Basti}, {Bawaj},
  {Bayley}, {Baylor}, {Bazzan}, {B{\'e}csy}, {Bedakihale}, {Bejger},
  {Belahcene}, {Benedetto}, {Beniwal}, {Bennett}, {Bentley}, {Benyaala},
  {Bergamin}, {Berger}, {Bernuzzi}, {Berry}, {Bersanetti}, {Bertolini},
  {Betzwieser}, {Beveridge}, {Bhandare}, {Bhardwaj}, {Bhattacharjee},
  {Bhaumik}, {Bilenko}, {Billingsley}, {Bini}, {Birney}, {Birnholtz},
  {Biscans}, {Bischi}, {Biscoveanu}, {Bisht}, {Biswas}, {Bitossi}, {Bizouard},
  {Blackburn}, {Blair}, {Blair}, {Blair}, {Bobba}, {Bode}, {Boer}, {Bogaert},
  {Boldrini}, {Bonavena}, {Bondu}, {Bonilla}, {Bonnand}, {Booker}, {Boom},
  {Bork}, {Boschi}, {Bose}, {Bose}, {Bossilkov}, {Boudart}, {Bouffanais},
  {Bozzi}, {Bradaschia}, {Brady}, {Bramley}, {Branch}, {Branchesi}, {Brandt},
  {Brau}, {Breschi}, {Briant}, {Briggs}, {Brillet}, {Brinkmann}, {Brockill},
  {Brooks}, {Brooks}, {Brown}, {Brunett}, {Bruno}, {Bruntz}, {Bryant}, {Bulik},
  {Bulten}, {Buonanno}, {Buscicchio}, {Buskulic}, {Buy}, {Byer}, {Cadonati},
  {Cagnoli}, {Cahillane}, {Bustillo}, {Callaghan}, {Callister}, {Calloni},
  {Cameron}, {Camp}, {Canepa}, {Canevarolo}, {Cannavacciuolo}, {Cannon}, {Cao},
  {Cao}, {Capocasa}, {Capote}, \& {Carapella}}]{2023PhRvX..13a1048A}
---. 2023{\natexlab{b}}, Physical Review X, 13, 011048,
  \dodoi{10.1103/PhysRevX.13.011048}

\bibitem[{{Agazie} {et~al.}(2023{\natexlab{a}}){Agazie}, {Anumarlapudi},
  {Archibald}, {Arzoumanian}, {Baker}, {B{\'e}csy}, {Blecha}, {Brazier},
  {Brook}, {Burke-Spolaor}, {Burnette}, {Case}, {Charisi}, {Chatterjee},
  {Chatziioannou}, {Cheeseboro}, {Chen}, {Cohen}, {Cordes}, {Cornish},
  {Crawford}, {Cromartie}, {Crowter}, {Cutler}, {Decesar}, {Degan}, {Demorest},
  {Deng}, {Dolch}, {Drachler}, {Ellis}, {Ferrara}, {Fiore}, {Fonseca},
  {Freedman}, {Garver-Daniels}, {Gentile}, {Gersbach}, {Glaser}, {Good},
  {G{\"u}ltekin}, {Hazboun}, {Hourihane}, {Islo}, {Jennings}, {Johnson},
  {Jones}, {Kaiser}, {Kaplan}, {Kelley}, {Kerr}, {Key}, {Klein}, {Laal}, {Lam},
  {Lamb}, {Lazio}, {Lewandowska}, {Littenberg}, {Liu}, {Lommen}, {Lorimer},
  {Luo}, {Lynch}, {Ma}, {Madison}, {Mattson}, {McEwen}, {McKee}, {McLaughlin},
  {McMann}, {Meyers}, {Meyers}, {Mingarelli}, {Mitridate}, {Natarajan}, {Ng},
  {Nice}, {Ocker}, {Olum}, {Pennucci}, {Perera}, {Petrov}, {Pol}, {Radovan},
  {Ransom}, {Ray}, {Romano}, {Sardesai}, {Schmiedekamp}, {Schmiedekamp},
  {Schmitz}, {Schult}, {Shapiro-Albert}, {Siemens}, {Simon}, {Siwek}, {Stairs},
  {Stinebring}, {Stovall}, {Sun}, {Susobhanan}, {Swiggum}, {Taylor}, {Taylor},
  {Turner}, {Unal}, {Vallisneri}, {van Haasteren}, {Vigeland}, {Wahl}, {Wang},
  {Witt}, {Young}, \& {Nanograv Collaboration}}]{2023ApJ...951L...8A}
{Agazie}, G., {Anumarlapudi}, A., {Archibald}, A.~M., {et~al.}
  2023{\natexlab{a}}, \apjl, 951, L8, \dodoi{10.3847/2041-8213/acdac6}

\bibitem[{{Agazie} {et~al.}(2023{\natexlab{b}}){Agazie}, {Anumarlapudi},
  {Archibald}, {Baker}, {B{\'e}csy}, {Blecha}, {Bonilla}, {Brazier}, {Brook},
  {Burke-Spolaor}, {Burnette}, {Case}, {Casey-Clyde}, {Charisi}, {Chatterjee},
  {Chatziioannou}, {Cheeseboro}, {Chen}, {Cohen}, {Cordes}, {Cornish},
  {Crawford}, {Cromartie}, {Crowter}, {Cutler}, {D'Orazio}, {Decesar}, {Degan},
  {Demorest}, {Deng}, {Dolch}, {Drachler}, {Ferrara}, {Fiore}, {Fonseca},
  {Freedman}, {Gardiner}, {Garver-Daniels}, {Gentile}, {Gersbach}, {Glaser},
  {Good}, {G{\"u}ltekin}, {Hazboun}, {Hourihane}, {Islo}, {Jennings},
  {Johnson}, {Jones}, {Kaiser}, {Kaplan}, {Kelley}, {Kerr}, {Key}, {Laal},
  {Lam}, {Lamb}, {Lazio}, {Lewandowska}, {Littenberg}, {Liu}, {Luo}, {Lynch},
  {Ma}, {Madison}, {McEwen}, {McKee}, {McLaughlin}, {McMann}, {Meyers},
  {Meyers}, {Mingarelli}, {Mitridate}, {Natarajan}, {Ng}, {Nice}, {Ocker},
  {Olum}, {Pennucci}, {Perera}, {Petrov}, {Pol}, {Radovan}, {Ransom}, {Ray},
  {Romano}, {Runnoe}, {Sardesai}, {Schmiedekamp}, {Schmiedekamp}, {Schmitz},
  {Schult}, {Shapiro-Albert}, {Siemens}, {Simon}, {Siwek}, {Stairs},
  {Stinebring}, {Stovall}, {Sun}, {Susobhanan}, {Swiggum}, {Taylor}, {Taylor},
  {Turner}, {Unal}, {Vallisneri}, {Vigeland}, {Wachter}, {Wahl}, {Wang},
  {Witt}, {Wright}, {Young}, \& {Nanograv Collaboration}}]{2023ApJ...952L..37A}
---. 2023{\natexlab{b}}, \apjl, 952, L37, \dodoi{10.3847/2041-8213/ace18b}

\bibitem[{{Agazie} {et~al.}(2023{\natexlab{c}}){Agazie}, {Anumarlapudi},
  {Archibald}, {Arzoumanian}, {Baker}, {B{\'e}csy}, {Blecha}, {Brazier},
  {Brook}, {Burke-Spolaor}, {Casey-Clyde}, {Charisi}, {Chatterjee}, {Cohen},
  {Cordes}, {Cornish}, {Crawford}, {Cromartie}, {Crowter}, {DeCesar},
  {Demorest}, {Dolch}, {Drachler}, {Ferrara}, {Fiore}, {Fonseca}, {Freedman},
  {Gardiner}, {Garver-Daniels}, {Gentile}, {Glaser}, {Good}, {G{\"u}ltekin},
  {Hazboun}, {Jennings}, {Johnson}, {Jones}, {Kaiser}, {Kaplan}, {Kelley},
  {Kerr}, {Key}, {Laal}, {Lam}, {Lamb}, {Lazio}, {Lewandowska}, {Liu},
  {Lorimer}, {Luo}, {Lynch}, {Ma}, {Madison}, {McEwen}, {McKee}, {McLaughlin},
  {McMann}, {Meyers}, {Mingarelli}, {Mitridate}, {Ng}, {Nice}, {Ocker}, {Olum},
  {Pennucci}, {Perera}, {Pol}, {Radovan}, {Ransom}, {Ray}, {Romano},
  {Sardesai}, {Schmiedekamp}, {Schmiedekamp}, {Schmitz}, {Schult},
  {Shapiro-Albert}, {Siemens}, {Simon}, {Siwek}, {Stairs}, {Stinebring},
  {Stovall}, {Susobhanan}, {Swiggum}, {Taylor}, {Turner}, {Unal}, {Vallisneri},
  {Vigeland}, {Wahl}, {Witt}, \& {Young}}]{2023ApJ...956L...3A}
---. 2023{\natexlab{c}}, \apjl, 956, L3, \dodoi{10.3847/2041-8213/acf4fd}

\bibitem[{{Ajith} {et~al.}(2008){Ajith}, {Babak}, {Chen}, {Hewitson},
  {Krishnan}, {Sintes}, {Whelan}, {Br{\"u}gmann}, {Diener}, {Dorband},
  {Gonzalez}, {Hannam}, {Husa}, {Pollney}, {Rezzolla}, {Santamar{\'\i}a},
  {Sperhake}, \& {Thornburg}}]{2008PhRvD..77j4017A}
{Ajith}, P., {Babak}, S., {Chen}, Y., {et~al.} 2008, \prd, 77, 104017,
  \dodoi{10.1103/PhysRevD.77.104017}

\bibitem[{{Alonso} {et~al.}(2020{\natexlab{a}}){Alonso}, {Contaldi}, {Cusin},
  {Ferreira}, \& {Renzini}}]{2020PhRvD.101l4048A}
{Alonso}, D., {Contaldi}, C.~R., {Cusin}, G., {Ferreira}, P.~G., \& {Renzini},
  A.~I. 2020{\natexlab{a}}, \prd, 101, 124048,
  \dodoi{10.1103/PhysRevD.101.124048}

\bibitem[{{Alonso} {et~al.}(2020{\natexlab{b}}){Alonso}, {Cusin}, {Ferreira},
  \& {Pitrou}}]{2020PhRvD.102b3002A}
{Alonso}, D., {Cusin}, G., {Ferreira}, P.~G., \& {Pitrou}, C.
  2020{\natexlab{b}}, \prd, 102, 023002, \dodoi{10.1103/PhysRevD.102.023002}

\bibitem[{{Amaro-Seoane} {et~al.}(2017){Amaro-Seoane}, {Audley}, {Babak},
  {Baker}, {Barausse}, {Bender}, {Berti}, {Binetruy}, {Born}, {Bortoluzzi},
  {Camp}, {Caprini}, {Cardoso}, {Colpi}, {Conklin}, {Cornish}, {Cutler},
  {Danzmann}, {Dolesi}, {Ferraioli}, {Ferroni}, {Fitzsimons}, {Gair}, {Gesa
  Bote}, {Giardini}, {Gibert}, {Grimani}, {Halloin}, {Heinzel}, {Hertog},
  {Hewitson}, {Holley-Bockelmann}, {Hollington}, {Hueller}, {Inchauspe},
  {Jetzer}, {Karnesis}, {Killow}, {Klein}, {Klipstein}, {Korsakova}, {Larson},
  {Livas}, {Lloro}, {Man}, {Mance}, {Martino}, {Mateos}, {McKenzie},
  {McWilliams}, {Miller}, {Mueller}, {Nardini}, {Nelemans}, {Nofrarias},
  {Petiteau}, {Pivato}, {Plagnol}, {Porter}, {Reiche}, {Robertson},
  {Robertson}, {Rossi}, {Russano}, {Schutz}, {Sesana}, {Shoemaker}, {Slutsky},
  {Sopuerta}, {Sumner}, {Tamanini}, {Thorpe}, {Troebs}, {Vallisneri},
  {Vecchio}, {Vetrugno}, {Vitale}, {Volonteri}, {Wanner}, {Ward}, {Wass},
  {Weber}, {Ziemer}, \& {Zweifel}}]{2017arXiv170200786A}
{Amaro-Seoane}, P., {Audley}, H., {Babak}, S., {et~al.} 2017, arXiv e-prints,
  arXiv:1702.00786, \dodoi{10.48550/arXiv.1702.00786}

\bibitem[{{Astropy Collaboration} {et~al.}(2013){Astropy Collaboration},
  {Robitaille}, {Tollerud}, {Greenfield}, {Droettboom}, {Bray}, {Aldcroft},
  {Davis}, {Ginsburg}, {Price-Whelan}, {Kerzendorf}, {Conley}, {Crighton},
  {Barbary}, {Muna}, {Ferguson}, {Grollier}, {Parikh}, {Nair}, {Unther},
  {Deil}, {Woillez}, {Conseil}, {Kramer}, {Turner}, {Singer}, {Fox}, {Weaver},
  {Zabalza}, {Edwards}, {Azalee Bostroem}, {Burke}, {Casey}, {Crawford},
  {Dencheva}, {Ely}, {Jenness}, {Labrie}, {Lian Lim}, {Pierfederici},
  {Pontzen}, {Ptak}, {Refsdal}, {Servillat}, \& {Streicher}}]{astropy:2013}
{Astropy Collaboration}, {Robitaille}, T.~P., {Tollerud}, E.~J., {et~al.} 2013,
  \aap, 558, A33, \dodoi{10.1051/0004-6361/201322068}

\bibitem[{{Astropy Collaboration} {et~al.}(2018){Astropy Collaboration},
  {Price-Whelan}, {Sip{\H o}cz}, {G{\"u}nther}, {Lim}, {Crawford}, {Conseil},
  {Shupe}, {Craig}, {Dencheva}, {Ginsburg}, {VanderPlas}, {Bradley},
  {P{\'e}rez-Su{\'a}rez}, {de Val-Borro}, {Paper Contributors}, {Aldcroft},
  {Cruz}, {Robitaille}, {Tollerud}, {Coordination Committee}, {Ardelean},
  {Babej}, {Bach}, {Bachetti}, {Bakanov}, {Bamford}, {Barentsen}, {Barmby},
  {Baumbach}, {Berry}, {Biscani}, {Boquien}, {Bostroem}, {Bouma}, {Brammer},
  {Bray}, {Breytenbach}, {Buddelmeijer}, {Burke}, {Calderone}, {Cano
  Rodr{\'{\i}}guez}, {Cara}, {Cardoso}, {Cheedella}, {Copin}, {Corrales},
  {Crichton}, {D{\'A}vella}, {Deil}, {Depagne}, {Dietrich}, {Donath},
  {Droettboom}, {Earl}, {Erben}, {Fabbro}, {Ferreira}, {Finethy}, {Fox},
  {Garrison}, {Gibbons}, {Goldstein}, {Gommers}, {Greco}, {Greenfield},
  {Groener}, {Grollier}, {Hagen}, {Hirst}, {Homeier}, {Horton}, {Hosseinzadeh},
  {Hu}, {Hunkeler}, {Ivezi{\'c}}, {Jain}, {Jenness}, {Kanarek}, {Kendrew},
  {Kern}, {Kerzendorf}, {Khvalko}, {King}, {Kirkby}, {Kulkarni}, {Kumar},
  {Lee}, {Lenz}, {Littlefair}, {Ma}, {Macleod}, {Mastropietro}, {McCully},
  {Montagnac}, {Morris}, {Mueller}, {Mumford}, {Muna}, {Murphy}, {Nelson},
  {Nguyen}, {Ninan}, {N{\"o}the}, {Ogaz}, {Oh}, {Parejko}, {Parley}, {Pascual},
  {Patil}, {Patil}, {Plunkett}, {Prochaska}, {Rastogi}, {Reddy Janga},
  {Sabater}, {Sakurikar}, {Seifert}, {Sherbert}, {Sherwood-Taylor}, {Shih},
  {Sick}, {Silbiger}, {Singanamalla}, {Singer}, {Sladen}, {Sooley},
  {Sornarajah}, {Streicher}, {Teuben}, {Thomas}, {Tremblay}, {Turner},
  {Terr{\'o}n}, {van Kerkwijk}, {de la Vega}, {Watkins}, {Weaver}, {Whitmore},
  {Woillez}, {Zabalza}, \& {Contributors}}]{astropy:2018}
{Astropy Collaboration}, {Price-Whelan}, A.~M., {Sip{\H o}cz}, B.~M., {et~al.}
  2018, \aj, 156, 123, \dodoi{10.3847/1538-3881/aabc4f}

\bibitem[{{Astropy Collaboration} {et~al.}(2022){Astropy Collaboration},
  {Price-Whelan}, {Lim}, {Earl}, {Starkman}, {Bradley}, {Shupe}, {Patil},
  {Corrales}, {Brasseur}, {N{\"o}the}, {Donath}, {Tollerud}, {Morris},
  {Ginsburg}, {Vaher}, {Weaver}, {Tocknell}, {Jamieson}, {van Kerkwijk},
  {Robitaille}, {Merry}, {Bachetti}, {G{\"u}nther}, {Aldcroft},
  {Alvarado-Montes}, {Archibald}, {B{\'o}di}, {Bapat}, {Barentsen},
  {Baz{\'a}n}, {Biswas}, {Boquien}, {Burke}, {Cara}, {Cara}, {Conroy},
  {Conseil}, {Craig}, {Cross}, {Cruz}, {D'Eugenio}, {Dencheva}, {Devillepoix},
  {Dietrich}, {Eigenbrot}, {Erben}, {Ferreira}, {Foreman-Mackey}, {Fox},
  {Freij}, {Garg}, {Geda}, {Glattly}, {Gondhalekar}, {Gordon}, {Grant},
  {Greenfield}, {Groener}, {Guest}, {Gurovich}, {Handberg}, {Hart},
  {Hatfield-Dodds}, {Homeier}, {Hosseinzadeh}, {Jenness}, {Jones}, {Joseph},
  {Kalmbach}, {Karamehmetoglu}, {Ka{\l}uszy{\'n}ski}, {Kelley}, {Kern},
  {Kerzendorf}, {Koch}, {Kulumani}, {Lee}, {Ly}, {Ma}, {MacBride}, {Maljaars},
  {Muna}, {Murphy}, {Norman}, {O'Steen}, {Oman}, {Pacifici}, {Pascual},
  {Pascual-Granado}, {Patil}, {Perren}, {Pickering}, {Rastogi}, {Roulston},
  {Ryan}, {Rykoff}, {Sabater}, {Sakurikar}, {Salgado}, {Sanghi}, {Saunders},
  {Savchenko}, {Schwardt}, {Seifert-Eckert}, {Shih}, {Jain}, {Shukla}, {Sick},
  {Simpson}, {Singanamalla}, {Singer}, {Singhal}, {Sinha}, {Sip{\H{o}}cz},
  {Spitler}, {Stansby}, {Streicher}, {{\v{S}}umak}, {Swinbank}, {Taranu},
  {Tewary}, {Tremblay}, {Val-Borro}, {Van Kooten}, {Vasovi{\'c}}, {Verma}, {de
  Miranda Cardoso}, {Williams}, {Wilson}, {Winkel}, {Wood-Vasey}, {Xue},
  {Yoachim}, {Zhang}, {Zonca}, \& {Astropy Project
  Contributors}}]{astropy:2022}
{Astropy Collaboration}, {Price-Whelan}, A.~M., {Lim}, P.~L., {et~al.} 2022,
  \apj, 935, 167, \dodoi{10.3847/1538-4357/ac7c74}

\bibitem[{{Bavera} {et~al.}(2022){Bavera}, {Franciolini}, {Cusin}, {Riotto},
  {Zevin}, \& {Fragos}}]{2022A&A...660A..26B}
{Bavera}, S.~S., {Franciolini}, G., {Cusin}, G., {et~al.} 2022, \aap, 660, A26,
  \dodoi{10.1051/0004-6361/202142208}

\bibitem[{{B{\'e}csy} {et~al.}(2022){B{\'e}csy}, {Cornish}, \&
  {Kelley}}]{2022ApJ...941..119B}
{B{\'e}csy}, B., {Cornish}, N.~J., \& {Kelley}, L.~Z. 2022, \apj, 941, 119,
  \dodoi{10.3847/1538-4357/aca1b2}

\bibitem[{{Bertacca} {et~al.}(2020){Bertacca}, {Ricciardone}, {Bellomo},
  {Jenkins}, {Matarrese}, {Raccanelli}, {Regimbau}, \&
  {Sakellariadou}}]{2020PhRvD.101j3513B}
{Bertacca}, D., {Ricciardone}, A., {Bellomo}, N., {et~al.} 2020, \prd, 101,
  103513, \dodoi{10.1103/PhysRevD.101.103513}

\bibitem[{{Blaizot} {et~al.}(2005){Blaizot}, {Wadadekar}, {Guiderdoni},
  {Colombi}, {Bertin}, {Bouchet}, {Devriendt}, \&
  {Hatton}}]{2005MNRAS.360..159B}
{Blaizot}, J., {Wadadekar}, Y., {Guiderdoni}, B., {et~al.} 2005, \mnras, 360,
  159, \dodoi{10.1111/j.1365-2966.2005.09019.x}

\bibitem[{{Breivik} {et~al.}(2020{\natexlab{a}}){Breivik}, Coughlin, Zevin,
  Rodriguez, Andrews, Kimball, mcdigman, \&
  1nhtran}]{katie_breivik_2020_3905335}
{Breivik}, K., Coughlin, S., Zevin, M., {et~al.} 2020{\natexlab{a}}, Zenodo,
  COSMIC v3.3.0, \dodoi{10.5281/zenodo.3905335}

\bibitem[{{Breivik} {et~al.}(2020{\natexlab{b}}){Breivik}, {Coughlin}, {Zevin},
  {Rodriguez}, {Kremer}, {Ye}, {Andrews}, {Kurkowski}, {Digman}, {Larson}, \&
  {Rasio}}]{2020ApJ...898...71B}
{Breivik}, K., {Coughlin}, S., {Zevin}, M., {et~al.} 2020{\natexlab{b}}, \apj,
  898, 71, \dodoi{10.3847/1538-4357/ab9d85}

\bibitem[{{Capurri} {et~al.}(2021){Capurri}, {Lapi}, {Baccigalupi}, {Boco},
  {Scelfo}, \& {Ronconi}}]{2021JCAP...11..032C}
{Capurri}, G., {Lapi}, A., {Baccigalupi}, C., {et~al.} 2021, \jcap, 2021, 032,
  \dodoi{10.1088/1475-7516/2021/11/032}

\bibitem[{{Christensen}(2019)}]{2019RPPh...82a6903C}
{Christensen}, N. 2019, Reports on Progress in Physics, 82, 016903,
  \dodoi{10.1088/1361-6633/aae6b5}

\bibitem[{{Cusin} {et~al.}(2018){Cusin}, {Dvorkin}, {Pitrou}, \&
  {Uzan}}]{2018PhRvL.120w1101C}
{Cusin}, G., {Dvorkin}, I., {Pitrou}, C., \& {Uzan}, J.-P. 2018, \prl, 120,
  231101, \dodoi{10.1103/PhysRevLett.120.231101}

\bibitem[{{Cusin} {et~al.}(2019){Cusin}, {Dvorkin}, {Pitrou}, \&
  {Uzan}}]{2019PhRvD.100f3004C}
---. 2019, \prd, 100, 063004, \dodoi{10.1103/PhysRevD.100.063004}

\bibitem[{{Cusin} {et~al.}(2017){Cusin}, {Pitrou}, \&
  {Uzan}}]{2017PhRvD..96j3019C}
{Cusin}, G., {Pitrou}, C., \& {Uzan}, J.-P. 2017, \prd, 96, 103019,
  \dodoi{10.1103/PhysRevD.96.103019}

\bibitem[{{De Lucia} \& {Blaizot}(2007)}]{2007MNRAS.375....2D}
{De Lucia}, G., \& {Blaizot}, J. 2007, \mnras, 375, 2,
  \dodoi{10.1111/j.1365-2966.2006.11287.x}

\bibitem[{{EPTA Collaboration} {et~al.}(2023){EPTA Collaboration}, {InPTA
  Collaboration}, {Antoniadis}, {Arumugam}, {Arumugam}, {Babak}, {Bagchi}, {Bak
  Nielsen}, {Bassa}, {Bathula}, {Berthereau}, {Bonetti}, {Bortolas}, {Brook},
  {Burgay}, {Caballero}, {Chalumeau}, {Champion}, {Chanlaridis}, {Chen},
  {Cognard}, {Dandapat}, {Deb}, {Desai}, {Desvignes}, {Dhanda-Batra},
  {Dwivedi}, {Falxa}, {Ferdman}, {Franchini}, {Gair}, {Goncharov}, {Gopakumar},
  {Graikou}, {Grie{\ss}meier}, {Guillemot}, {Guo}, {Gupta}, {Hisano}, {Hu},
  {Iraci}, {Izquierdo-Villalba}, {Jang}, {Jawor}, {Janssen}, {Jessner},
  {Joshi}, {Kareem}, {Karuppusamy}, {Keane}, {Keith}, {Kharbanda}, {Kikunaga},
  {Kolhe}, {Kramer}, {Krishnakumar}, {Lackeos}, {Lee}, {Liu}, {Liu}, {Lyne},
  {McKee}, {Maan}, {Main}, {Mickaliger}, {Ni{\c{t}}u}, {Nobleson}, {Paladi},
  {Parthasarathy}, {Perera}, {Perrodin}, {Petiteau}, {Porayko}, {Possenti},
  {Prabu}, {Quelquejay Leclere}, {Rana}, {Samajdar}, {Sanidas}, {Sesana},
  {Shaifullah}, {Singha}, {Speri}, {Spiewak}, {Srivastava}, {Stappers},
  {Surnis}, {Susarla}, {Susobhanan}, {Takahashi}, {Tarafdar}, {Theureau},
  {Tiburzi}, {van der Wateren}, {Vecchio}, {Venkatraman Krishnan}, {Verbiest},
  {Wang}, {Wang}, \& {Wu}}]{2023A&A...678A..50E}
{EPTA Collaboration}, {InPTA Collaboration}, {Antoniadis}, J., {et~al.} 2023,
  \aap, 678, A50, \dodoi{10.1051/0004-6361/202346844}

\bibitem[{{Gardiner} {et~al.}(2024){Gardiner}, {Kelley}, {Lemke}, \&
  {Mitridate}}]{2024ApJ...965..164G}
{Gardiner}, E.~C., {Kelley}, L.~Z., {Lemke}, A.-M., \& {Mitridate}, A. 2024,
  \apj, 965, 164, \dodoi{10.3847/1538-4357/ad2be8}

\bibitem[{{G{\'o}rski} {et~al.}(2005){G{\'o}rski}, {Hivon}, {Banday},
  {Wandelt}, {Hansen}, {Reinecke}, \& {Bartelmann}}]{2005ApJ...622..759G}
{G{\'o}rski}, K.~M., {Hivon}, E., {Banday}, A.~J., {et~al.} 2005, \apj, 622,
  759, \dodoi{10.1086/427976}

\bibitem[{G{\'o}rski {et~al.}(2024)G{\'o}rski, Wandelt, Hivon, Hansen, \&
  Banday}]{gorski2024healpixprimer}
G{\'o}rski, K.~M., Wandelt, B.~D., Hivon, E., Hansen, F.~K., \& Banday, A.~J.
  2024, The HEALPix Primer, v3.83.
\newblock \url{https://healpix.sourceforge.io/pdf/intro.pdf}

\bibitem[{{Guo} {et~al.}(2011){Guo}, {White}, {Boylan-Kolchin}, {De Lucia},
  {Kauffmann}, {Lemson}, {Li}, {Springel}, \& {Weinmann}}]{2011MNRAS.413..101G}
{Guo}, Q., {White}, S., {Boylan-Kolchin}, M., {et~al.} 2011, \mnras, 413, 101,
  \dodoi{10.1111/j.1365-2966.2010.18114.x}

\bibitem[{Harris {et~al.}(2020)Harris, Millman, van~der Walt, Gommers,
  Virtanen, Cournapeau, Wieser, Taylor, Berg, Smith, Kern, Picus, Hoyer, van
  Kerkwijk, Brett, Haldane, Fernández~del Río, Wiebe, Peterson,
  Gérard-Marchant, Sheppard, Reddy, Weckesser, Abbasi, Gohlke, \&
  Oliphant}]{2020NumPy-Array}
Harris, C.~R., Millman, K.~J., van~der Walt, S.~J., {et~al.} 2020, Nature, 585,
  357–362, \dodoi{10.1038/s41586-020-2649-2}

\bibitem[{{Harry} \& {LIGO Scientific
  Collaboration}(2010)}]{2010CQGra..27h4006H}
{Harry}, G.~M., \& {LIGO Scientific Collaboration}. 2010, Classical and Quantum
  Gravity, 27, 084006, \dodoi{10.1088/0264-9381/27/8/084006}

\bibitem[{Hunter(2007)}]{Hunter:2007}
Hunter, J.~D. 2007, Computing in Science \& Engineering, 9, 90,
  \dodoi{10.1109/MCSE.2007.55}

\bibitem[{{Jenkins}(2022)}]{2022arXiv220205105J}
{Jenkins}, A.~C. 2022, arXiv e-prints, arXiv:2202.05105,
  \dodoi{10.48550/arXiv.2202.05105}

\bibitem[{{Jenkins} {et~al.}(2019{\natexlab{a}}){Jenkins}, {O'Shaughnessy},
  {Sakellariadou}, \& {Wysocki}}]{2019PhRvL.122k1101J}
{Jenkins}, A.~C., {O'Shaughnessy}, R., {Sakellariadou}, M., \& {Wysocki}, D.
  2019{\natexlab{a}}, \prl, 122, 111101, \dodoi{10.1103/PhysRevLett.122.111101}

\bibitem[{{Jenkins} {et~al.}(2019{\natexlab{b}}){Jenkins}, {Romano}, \&
  {Sakellariadou}}]{2019PhRvD.100h3501J}
{Jenkins}, A.~C., {Romano}, J.~D., \& {Sakellariadou}, M. 2019{\natexlab{b}},
  \prd, 100, 083501, \dodoi{10.1103/PhysRevD.100.083501}

\bibitem[{{Jenkins} \& {Sakellariadou}(2018)}]{2018PhRvD..98f3509J}
{Jenkins}, A.~C., \& {Sakellariadou}, M. 2018, \prd, 98, 063509,
  \dodoi{10.1103/PhysRevD.98.063509}

\bibitem[{{Jenkins} \& {Sakellariadou}(2019)}]{2019PhRvD.100f3508J}
---. 2019, \prd, 100, 063508, \dodoi{10.1103/PhysRevD.100.063508}

\bibitem[{{Jenkins} {et~al.}(2018){Jenkins}, {Sakellariadou}, {Regimbau}, \&
  {Slezak}}]{2018PhRvD..98f3501J}
{Jenkins}, A.~C., {Sakellariadou}, M., {Regimbau}, T., \& {Slezak}, E. 2018,
  \prd, 98, 063501, \dodoi{10.1103/PhysRevD.98.063501}

\bibitem[{{Jiang} {et~al.}(2019){Jiang}, {Wang}, {Gao}, {Zhang}, {Guo}, {Wang},
  \& {Pan}}]{2019RAA....19..151J}
{Jiang}, Z., {Wang}, J., {Gao}, L., {et~al.} 2019, Research in Astronomy and
  Astrophysics, 19, 151, \dodoi{10.1088/1674-4527/19/10/151}

\bibitem[{{Kitzbichler} \& {White}(2007)}]{2007MNRAS.376....2K}
{Kitzbichler}, M.~G., \& {White}, S.~D.~M. 2007, \mnras, 376, 2,
  \dodoi{10.1111/j.1365-2966.2007.11458.x}

\bibitem[{{Kluyver} {et~al.}(2016){Kluyver}, {Ragan-Kelley}, {P{\'e}rez},
  {Granger}, {Bussonnier}, {Frederic}, {Kelley}, {Hamrick}, {Grout}, {Corlay},
  {Ivanov}, {Avila}, {Abdalla}, {Willing}, \& {Jupyter Development
  Team}}]{2016ppap.book...87K}
{Kluyver}, T., {Ragan-Kelley}, B., {P{\'e}rez}, F., {et~al.} 2016, in IOS
  Press, 87--90, \dodoi{10.3233/978-1-61499-649-1-87}

\bibitem[{{Korytov} {et~al.}(2019){Korytov}, {Hearin}, {Kovacs}, {Larsen},
  {Rangel}, {Hollowed}, {Benson}, {Heitmann}, {Mao}, {Bahmanyar}, {Chang},
  {Campbell}, {DeRose}, {Finkel}, {Frontiere}, {Gawiser}, {Habib}, {Joachimi},
  {Lanusse}, {Li}, {Mandelbaum}, {Morrison}, {Newman}, {Pope}, {Rykoff},
  {Simet}, {To}, {Vikraman}, {Wechsler}, {White}, \& {(The LSST Dark Energy
  Science Collaboration}}]{2019ApJS..245...26K}
{Korytov}, D., {Hearin}, A., {Kovacs}, E., {et~al.} 2019, \apjs, 245, 26,
  \dodoi{10.3847/1538-4365/ab510c}

\bibitem[{{Kouvatsos} {et~al.}(2024){Kouvatsos}, {Jenkins}, {Renzini},
  {Romano}, \& {Sakellariadou}}]{2024PhRvD.109j3535K}
{Kouvatsos}, N., {Jenkins}, A.~C., {Renzini}, A.~I., {Romano}, J.~D., \&
  {Sakellariadou}, M. 2024, \prd, 109, 103535,
  \dodoi{10.1103/PhysRevD.109.103535}

\bibitem[{{Lam} {et~al.}(2015){Lam}, {Pitrou}, \&
  {Seibert}}]{2015llvm.confE...1L}
{Lam}, S.~K., {Pitrou}, A., \& {Seibert}, S. 2015, in Proc. Second Workshop on
  the LLVM Compiler Infrastructure in HPC, 1--6,
  \dodoi{10.1145/2833157.2833162}

\bibitem[{{Li} {et~al.}(2024){Li}, {Jiang}, {Fan}, {Chen}, {Gao}, {Guo}, \&
  {Yu}}]{2024MNRAS.527.5616L}
{Li}, Z., {Jiang}, Z., {Fan}, X.-L., {et~al.} 2024, \mnras, 527, 5616,
  \dodoi{10.1093/mnras/stad3576}

\bibitem[{{Merson} {et~al.}(2013){Merson}, {Baugh}, {Helly}, {Gonzalez-Perez},
  {Cole}, {Bielby}, {Norberg}, {Frenk}, {Benson}, {Bower}, {Lacey}, \&
  {Lagos}}]{2013MNRAS.429..556M}
{Merson}, A.~I., {Baugh}, C.~M., {Helly}, J.~C., {et~al.} 2013, \mnras, 429,
  556, \dodoi{10.1093/mnras/sts355}

\bibitem[{{Mingarelli} {et~al.}(2017){Mingarelli}, {Lazio}, {Sesana}, {Greene},
  {Ellis}, {Ma}, {Croft}, {Burke-Spolaor}, \& {Taylor}}]{2017NatAs...1..886M}
{Mingarelli}, C. M.~F., {Lazio}, T. J.~W., {Sesana}, A., {et~al.} 2017, Nature
  Astronomy, 1, 886, \dodoi{10.1038/s41550-017-0299-6}

\bibitem[{{Payne} {et~al.}(2020){Payne}, {Banagiri}, {Lasky}, \&
  {Thrane}}]{2020PhRvD.102j2004P}
{Payne}, E., {Banagiri}, S., {Lasky}, P.~D., \& {Thrane}, E. 2020, \prd, 102,
  102004, \dodoi{10.1103/PhysRevD.102.102004}

\bibitem[{{Phinney}(2001)}]{2001astro.ph..8028P}
{Phinney}, E.~S. 2001, arXiv e-prints, astro,
  \dodoi{10.48550/arXiv.astro-ph/0108028}

\bibitem[{{Punturo} {et~al.}(2010){Punturo}, {Abernathy}, {Acernese}, {Allen},
  {Andersson}, {Arun}, {Barone}, {Barr}, {Barsuglia}, {Beker}, {Beveridge},
  {Birindelli}, {Bose}, {Bosi}, {Braccini}, {Bradaschia}, {Bulik}, {Calloni},
  {Cella}, {Chassande Mottin}, {Chelkowski}, {Chincarini}, {Clark}, {Coccia},
  {Colacino}, {Colas}, {Cumming}, {Cunningham}, {Cuoco}, {Danilishin},
  {Danzmann}, {De Luca}, {De Salvo}, {Dent}, {De Rosa}, {Di Fiore}, {Di
  Virgilio}, {Doets}, {Fafone}, {Falferi}, {Flaminio}, {Franc}, {Frasconi},
  {Freise}, {Fulda}, {Gair}, {Gemme}, {Gennai}, {Giazotto}, {Glampedakis},
  {Granata}, {Grote}, {Guidi}, {Hammond}, {Hannam}, {Harms}, {Heinert},
  {Hendry}, {Heng}, {Hennes}, {Hild}, {Hough}, {Husa}, {Huttner}, {Jones},
  {Khalili}, {Kokeyama}, {Kokkotas}, {Krishnan}, {Lorenzini}, {L{\"u}ck},
  {Majorana}, {Mandel}, {Mandic}, {Martin}, {Michel}, {Minenkov}, {Morgado},
  {Mosca}, {Mours}, {M{\"u}ller{\textendash}Ebhardt}, {Murray}, {Nawrodt},
  {Nelson}, {Oshaughnessy}, {Ott}, {Palomba}, {Paoli}, {Parguez},
  {Pasqualetti}, {Passaquieti}, {Passuello}, {Pinard}, {Poggiani}, {Popolizio},
  {Prato}, {Puppo}, {Rabeling}, {Rapagnani}, {Read}, {Regimbau}, {Rehbein},
  {Reid}, {Rezzolla}, {Ricci}, {Richard}, {Rocchi}, {Rowan}, {R{\"u}diger},
  {Sassolas}, {Sathyaprakash}, {Schnabel}, {Schwarz}, {Seidel}, {Sintes},
  {Somiya}, {Speirits}, {Strain}, {Strigin}, {Sutton}, {Tarabrin},
  {Th{\"u}ring}, {van den Brand}, {van Leewen}, {van Veggel}, {van den Broeck},
  {Vecchio}, {Veitch}, {Vetrano}, {Vicere}, {Vyatchanin}, {Willke}, {Woan},
  {Wolfango}, \& {Yamamoto}}]{2010CQGra..27s4002P}
{Punturo}, M., {Abernathy}, M., {Acernese}, F., {et~al.} 2010, Classical and
  Quantum Gravity, 27, 194002, \dodoi{10.1088/0264-9381/27/19/194002}

\bibitem[{{Reardon} {et~al.}(2023){Reardon}, {Zic}, {Shannon}, {Hobbs},
  {Bailes}, {Di Marco}, {Kapur}, {Rogers}, {Thrane}, {Askew}, {Bhat},
  {Cameron}, {Cury{\l}o}, {Coles}, {Dai}, {Goncharov}, {Kerr}, {Kulkarni},
  {Levin}, {Lower}, {Manchester}, {Mandow}, {Miles}, {Nathan}, {Os{\l}owski},
  {Russell}, {Spiewak}, {Zhang}, \& {Zhu}}]{2023ApJ...951L...6R}
{Reardon}, D.~J., {Zic}, A., {Shannon}, R.~M., {et~al.} 2023, \apjl, 951, L6,
  \dodoi{10.3847/2041-8213/acdd02}

\bibitem[{{Renzini} {et~al.}(2022){Renzini}, {Goncharov}, {Jenkins}, \&
  {Meyers}}]{2022Galax..10...34R}
{Renzini}, A.~I., {Goncharov}, B., {Jenkins}, A.~C., \& {Meyers}, P.~M. 2022,
  Galaxies, 10, 34, \dodoi{10.3390/galaxies10010034}

\bibitem[{{Smith} {et~al.}(2022){Smith}, {Cole}, {Grove}, {Norberg}, \&
  {Zarrouk}}]{2022MNRAS.516.4529S}
{Smith}, A., {Cole}, S., {Grove}, C., {Norberg}, P., \& {Zarrouk}, P. 2022,
  \mnras, 516, 4529, \dodoi{10.1093/mnras/stac2519}

\bibitem[{{Spergel} {et~al.}(2003){Spergel}, {Verde}, {Peiris}, {Komatsu},
  {Nolta}, {Bennett}, {Halpern}, {Hinshaw}, {Jarosik}, {Kogut}, {Limon},
  {Meyer}, {Page}, {Tucker}, {Weiland}, {Wollack}, \&
  {Wright}}]{2003ApJS..148..175S}
{Spergel}, D.~N., {Verde}, L., {Peiris}, H.~V., {et~al.} 2003, \apjs, 148, 175,
  \dodoi{10.1086/377226}

\bibitem[{{Springel} {et~al.}(2005){Springel}, {White}, {Jenkins}, {Frenk},
  {Yoshida}, {Gao}, {Navarro}, {Thacker}, {Croton}, {Helly}, {Peacock}, {Cole},
  {Thomas}, {Couchman}, {Evrard}, {Colberg}, \& {Pearce}}]{2005Natur.435..629S}
{Springel}, V., {White}, S. D.~M., {Jenkins}, A., {et~al.} 2005, \nat, 435,
  629, \dodoi{10.1038/nature03597}

\bibitem[{Virtanen {et~al.}(2020)Virtanen, Gommers, Oliphant, Haberland, Reddy,
  Cournapeau, Burovski, Peterson, Weckesser, Bright, {van der Walt}, Brett,
  Wilson, Millman, Mayorov, Nelson, Jones, Kern, Larson, Carey, Polat, Feng,
  Moore, {VanderPlas}, Laxalde, Perktold, Cimrman, Henriksen, Quintero, Harris,
  Archibald, Ribeiro, Pedregosa, {van Mulbregt}, \& {SciPy 1.0
  Contributors}}]{2020SciPy-NMeth}
Virtanen, P., Gommers, R., Oliphant, T.~E., {et~al.} 2020, Nature Methods, 17,
  261, \dodoi{10.1038/s41592-019-0686-2}

\bibitem[{{Xu} {et~al.}(2023){Xu}, {Chen}, {Guo}, {Jiang}, {Wang}, {Xu}, {Xue},
  {Nicolas Caballero}, {Yuan}, {Xu}, {Wang}, {Hao}, {Luo}, {Lee}, {Han},
  {Jiang}, {Shen}, {Wang}, {Wang}, {Xu}, {Wu}, {Manchester}, {Qian}, {Guan},
  {Huang}, {Sun}, \& {Zhu}}]{2023RAA....23g5024X}
{Xu}, H., {Chen}, S., {Guo}, Y., {et~al.} 2023, Research in Astronomy and
  Astrophysics, 23, 075024, \dodoi{10.1088/1674-4527/acdfa5}

\bibitem[{{Yang} {et~al.}(2024){Yang}, {Guo}, {Cao}, {Shao}, \&
  {Yuan}}]{2024arXiv240805043Y}
{Yang}, Q., {Guo}, X., {Cao}, Z., {Shao}, X., \& {Yuan}, X. 2024, arXiv
  e-prints, arXiv:2408.05043, \dodoi{10.48550/arXiv.2408.05043}

\bibitem[{{Zhu} {et~al.}(2011){Zhu}, {Howell}, {Regimbau}, {Blair}, \&
  {Zhu}}]{2011ApJ...739...86Z}
{Zhu}, X.-J., {Howell}, E., {Regimbau}, T., {Blair}, D., \& {Zhu}, Z.-H. 2011,
  \apj, 739, 86, \dodoi{10.1088/0004-637X/739/2/86}

\bibitem[{Zonca {et~al.}(2019)Zonca, Singer, Lenz, Reinecke, Rosset, Hivon, \&
  Gorski}]{Zonca2019}
Zonca, A., Singer, L., Lenz, D., {et~al.} 2019, Journal of Open Source
  Software, 4, 1298, \dodoi{10.21105/joss.01298}

\end{thebibliography}
\bibliographystyle{aasjournal}
\end{document}